\documentclass[12pt,a4paper]{article}
	
\usepackage{a4wide}
\usepackage{amsmath}
\usepackage{amssymb}
\usepackage{amsfonts}
\usepackage{epsfig}
\usepackage{exscale}
\usepackage{float}
\usepackage{bbm}
\usepackage[numbers,sort&compress]{natbib}

\newcommand{\R}{{\mathbb{R}}}

\newcommand{\II}{\rm I\hspace{-0.2mm}I}
\newcommand{\I}{\rm I}
\newcommand{\p}{\partial}

\newcommand{\sign}{\text{sign}}

\makeatletter
\@addtoreset{equation}{section}
\makeatother

\title{Supersymmetric Descendants of Self-Adjointly Extended Quantum Mechanical
Hamiltonians}

\author{M.\ H.\ Al-Hashimi$^a$, M.\ Salman$^b$, A.\ Shalaby$^{b,c}$, and 
U.-J.\ Wiese$^{a,d}$ \\ \\
$^a$ Albert Einstein Center for Fundamental Physics \\
Institute for Theoretical Physics, Bern University \\
Sidlerstrasse 5, CH-3012 Bern, Switzerland \\ \\
$^b$ Department of Mathematics, Statistics, and Physics \\
Qatar University, Al Tarfa, Doha 2713, Qatar \\ \\
$^c$ Physics Department, Faculty of Science, Mansoura University, Egypt \\ \\
$^d$ Center for Theoretical Physics, Massachusetts Institute of Technology \\
77 Massachusetts Avenue, Cambridge, Massachusetts, U.S.A. \\ \\}

\begin{document} 

\maketitle

\vspace{-1cm}

\begin{abstract} \normalsize

We consider the descendants of self-adjointly extended Hamiltonians in 
supersymmetric quantum mechanics on a half-line, on an interval, and on a
punctured line or interval. While there is a 4-parameter family of 
self-adjointly extended Hamiltonians on a punctured line, only a 3-parameter 
sub-family has supersymmetric descendants that are themselves self-adjoint. We
also address the self-adjointness of an operator related to the supercharge, 
and point out that only a sub-class of its most general self-adjoint extensions 
is physical. Besides a general characterization of self-adjoint extensions 
and their supersymmetric descendants, we explicitly consider concrete examples, 
including a particle in a box with general boundary conditions, with and without
an additional point interaction. We also discuss bulk-boundary resonances and 
their manifestation in the supersymmetric descendant.

\end{abstract}

\newpage
 
\section{Introduction}

The differences between Hermiticity and self-adjointness of quantum mechanical 
operators \cite{Ree75} were first understood by von Neumann \cite{Neu32}, but 
are rarely emphasized in the quantum mechanics textbook literature. Even 
standard textbook problems such as a particle confined to a box \cite{Bon01} or 
endowed with a point interaction \cite{Ber61,Alb88,Jac91,Car90} become much 
richer when studied systematically in the context of self-adjointly extended 
Hamiltonians \cite{AlH12}. Self-adjoint extensions arise naturally at spatial 
boundaries, such as the interfaces in semiconductor heterostructures including 
quantum dots, quantum wires, and quantum wells \cite{Har05}, or at singular 
points, e.g.\ at the location of a cosmic string or vortex, which may manifest 
themselves as the tip of a cone in $(2+1)$ space-time dimensions 
\cite{Sou89,Sou89a,Sit96,AlH08}. At a spatial boundary, a real-valued 
self-adjoint extension parameter characterizes the so-called Robin boundary 
conditions, which interpolate between Dirichlet and Neumann boundary conditions 
\cite{Dan97,Are03,Are03a,Pan06,Bal70,Sch00}. Robin boundary conditions 
have also been used in quantum field theory, for example, in investigations of 
the Casimir effect \cite{Leb96,Alb04} and of the AdS/CFT correspondence 
\cite{Min00}. The confinement of atoms or 
molecules in a finite region of space is an important subject in nanotechnology 
\cite{Sab09}. Recently, we have derived a generalized Heisenberg uncertainty 
relation for a quantum dot with general Robin boundary conditions \cite{AlH12}. 
As special cases, we have investigated electrons in a spherical cavity bound to
its center by harmonic \cite{AlH13} or Coulomb forces \cite{AlH12a}, with a 
focus on the resulting accidental symmetries \cite{Pup98,Pup02}. General Robin 
boundary conditions may lead to bound states localized on the confining wall. 
In these cases, we have encountered bulk-boundary resonances whose wave 
functions are partly localized near the boundary and partly near the center of 
the cavity.

Supersymmetric quantum mechanics has deepened the understanding of quantum 
mechanics by associating a chain of supersymmetric descendants to a given 
quantum mechanical Hamiltonian \cite{Wit81,Coo83,Coo01,Bag01}. In this way, the 
number of analytically solvable quantum mechanical problems has been extended 
significantly. General point interactions have been investigated in the
framework of supersymmetric quantum mechanics in 
\cite{Che01,Uch03,Uch03a,Nag03,Nag04,Fal05}. Here we study the superpartners of 
self-adjointly extended Hamiltonians. Remarkably, these are not automatically 
self-adjoint. In particular, only a 3-parameter sub-family of the general 
4-parameter family of self-adjointly extended Hamiltonians on a punctured line 
have supersymmetric descendants that are themselves self-adjoint. We also 
construct the self-adjoint extensions of an operator related to the supercharge.
Since we consider a Hamiltonian and its supersymmetric descendant as two 
different physical systems, rather than as two parts of the same system, only a 
sub-class of self-adjoint extensions of this operator is physical.

We illustrate our general results
with specific systems, including a particle confined to a box with or without
an additional point interaction. We will again encounter bulk-boundary 
resonances, and we study their manifestation in the corresponding supersymmetric
descendant. While there are experimental realizations of bulk-boundary 
resonances, e.g.\ for atoms encapsulated in fullerenes \cite{Pus93,Wen93}, our 
current study is not motivated by a particular application. Instead, we aim at 
illuminating the relations between the theoretical concepts of supersymmetry 
and self-adjoint extensions in quantum mechanics in general. 

Some of the concrete systems studied here are so simple that they could easily 
serve as problems in the teaching of quantum mechanics. In this sense, our 
paper also has pedagogical intentions, by trying to convince the reader that 
the theory of self-adjoint extensions is not only mathematically elegant, but 
also of great physical relevance, that deserves a more prominent place in the
teaching of quantum mechanics.

The rest of this paper is organized as follows. In section 2 we consider the
general self-adjoint extensions of quantum mechanical Hamiltonians on a 
half-line as well as on a punctured line, and we address the issue of
self-adjointness of an operator constructed from the supercharge. In section 3, 
we investigate concrete examples of a particle confined to a box, or endowed 
with a point interaction, and we study the phenomenon of bulk-boundary 
resonances. Finally, section 4 contains our conclusions.

\section{Self-Adjointness of Supersymmetric Descendant Hamiltonians}

After briefly reviewing the basics of supersymmetric quantum mechanics
\cite{Wit81,Coo83}, in this section, we study the self-adjoint extensions of 
the Hamiltonian of a particle on a half-line and on a punctured line, and we 
investigate their supersymmetric descendants. In addition, we investigate the
self-adjointness of an operator constructed from the supercharge.

\subsection{Basics of Supersymmetric Quantum Mechanics}

In order to make this paper self-contained, let us briefly review the basics of
supersymmetric quantum mechanics in one spatial dimension. We consider a 
non-relativistic particle of mass $m$ moving under the influence of a potential
$V(x)$ such that
\begin{eqnarray}
&&- \frac{1}{2m} \p_x^2 \Psi_n(x) + V(x) \Psi_n(x) = E_n  \Psi_n(x) \
\Rightarrow \nonumber \\
&&- \p_x^2 \Psi_n(x) + 2 m V(x) \Psi_n(x) = \lambda_n  \Psi_n(x), \quad
\lambda_n = 2 m E_n.
\end{eqnarray}
Introducing the real-valued superpotential $W(x)$ we define
\begin{equation}
A = \p_x + W(x), \quad A^\dagger = - \p_x + W(x).
\end{equation}
We construct the Hamiltonian as
\begin{equation}
2 m H = A^\dagger A  = [- \p_x + W(x)][\p_x + W(x)] = 
- \p_x^2 - \p_x W(x) + W(x)^2,
\end{equation}
and thus identify the potential as
\begin{equation}
2 m V(x) = - \p_x W(x) + W(x)^2,
\end{equation}
such that
\begin{equation}
H \Psi_n(x) = E_n  \Psi_n(x) \ \Rightarrow \
2 m H \Psi_n(x) = \lambda_n  \Psi_n(x).
\end{equation}
Shifting the potential such that the ground state energy vanishes, 
i.e.\ $\lambda_0 = 0$, one obtains
\begin{equation}
H \Psi_0(x) = 0 \ \Rightarrow \ A \Psi_0(x) = [\p_x + W(x)] \Psi_0(x) = 0,
\end{equation}
which allows us to relate the superpotential to the ground state wave function
\begin{equation}
W(x) = - \frac{\p_x \Psi_0(x)}{\Psi_0(x)} = - \p_x \log\Psi_0(x).
\end{equation}

We now define the supersymmetric descendant Hamiltonian as
\begin{equation}
2 m H' = A A^\dagger = [\p_x + W(x)][- \p_x + W(x)] = 
- \p_x^2 + \p_x W(x) + W(x)^2,
\end{equation}
with the potential
\begin{equation}
2 m V'(x) = \p_x W(x) + W(x)^2.
\end{equation}
The eigenstates of $H'$ are related to the eigenstates of $H$ by
\begin{equation}
\label{astep}
\Psi_n'(x) = \frac{1}{\sqrt{\lambda_n}} A \Psi_n(x), \ n > 0, 
\end{equation}
such that indeed
\begin{equation}
2 m H' \Psi_n'(x) = \frac{1}{\sqrt{\lambda_n}} A A^\dagger A \Psi_n(x) = 
\sqrt{\lambda_n} A \Psi_n(x) = \lambda_n \Psi_n'(x).
\end{equation}
Hence, the spectrum of $H'$ coincides with the spectrum of excited states of 
$H$. The normalization of the superpartner wave functions is given by
\begin{equation}
\langle \Psi_n'|\Psi_n'\rangle = 
\frac{1}{\lambda_n} \langle \Psi_n|A^\dagger A|\Psi_n\rangle = 1.
\end{equation}
It should be noted that eq.(\ref{astep}) works even for the continuous part of
the spectrum, when the wave functions are not normalizable in the usual sense.
If the wave function 
\begin{equation}
\Psi_0(x) = C \exp\left(- \int_0^x dx' \ W(x')\right)
\end{equation}
is not square-integrable, $H$ does not have a normalizable ground state of 
zero energy. In that case, the spectra of $H$ and $H'$ are completely
identical and supersymmetry is spontaneously broken. In this paper we will only
encounter unbroken supersymmetry.

By applying the same procedure again, one obtains the next superpotential
\begin{equation}
W'(x) = - \frac{\p_x \Psi_1'(x)}{\Psi_1'(x)},
\end{equation}
such that
\begin{eqnarray}
&&2 m V'(x) - \lambda_1 = - \p_x W'(x) + W'(x)^2, \nonumber \\
&&2 m V''(x) - \lambda_1 = \p_x W'(x) + W'(x)^2.
\end{eqnarray}
By iterating the construction, as long as the system has a discrete spectrum of
bound states, one can then generate a chain of supersymmetric descendants.

Let us also consider the supercharge and its Hermitean conjugate
\begin{equation}
{\cal Q} = \left(\begin{array}{cc} 0 & 0 \\ A & 0 \end{array}\right), \quad
{\cal Q}^\dagger = 
\left(\begin{array}{cc} 0 & A^\dagger \\ 0 & 0 \end{array}\right),
\end{equation}
which generate the Hamiltonians $H$ and $H'$ by anti-commutation 
\begin{eqnarray}
\{{\cal Q},{\cal Q}^\dagger\}&=&
\left(\begin{array}{cc} 0 & 0 \\ A & 0 \end{array}\right)
\left(\begin{array}{cc} 0 & A^\dagger \\ 0 & 0 \end{array}\right) +
\left(\begin{array}{cc} 0 & A^\dagger \\ 0 & 0 \end{array}\right)
\left(\begin{array}{cc} 0 & 0 \\ A & 0 \end{array}\right) \nonumber \\
&=&\left(\begin{array}{cc} A^\dagger A & 0 \\ 0 & A A^\dagger \end{array}\right) =
\left(\begin{array}{cc} H & 0 \\ 0 & H' \end{array}\right) = {\cal H}.
\end{eqnarray}
The supercharge is nilpotent and commutes with the Hamiltonian, i.e.
\begin{equation}
{\cal Q}^2 = {{\cal Q}^\dagger}^2 = 0, \quad 
[{\cal Q},{\cal H}] = [{\cal Q}^\dagger,{\cal H}] = 0.
\end{equation}
While the supercharge itself is not Hermitean, it is natural to construct the 
operator
\begin{equation}
{\cal Q}_+ = {\cal Q} + {\cal Q}^\dagger = 
\left(\begin{array}{cc} 0 & A^\dagger \\ A & 0 \end{array}\right),
\end{equation}
which yields ${\cal Q}_+^2 = {\cal H}$, and is ``Hermitean'' at a rather formal 
level. Later, we'll properly address the self-adjointness of ${\cal Q}_+$ 
\cite{Fal05}.

\subsection{Particle on a Half-Line}

Let us consider a particle confined to a half-line, the region $x \leq x_0$.
If the region $x > x_0$ is energetically forbidden, the standard textbook 
approach is to demand Dirichlet boundary conditions, i.e.\ $\Psi(x_0) = 0$.
However, this is not really necessary. It suffices that the probability current
density,
\begin{equation}
j(x) = \frac{1}{2 m i} 
\left[\Psi(x)^* \p_x \Psi(x) - \p_x \Psi(x)^* \Psi(x)\right],
\end{equation}
vanishes at the boundary, i.e.\ $j(x) = 0$. This is the case, not only for the
Dirichlet boundary condition, but also for the more general Robin boundary 
condition
\begin{equation}
\label{bc}
\p_x \Psi(x_0) + \gamma \Psi(x_0) = 0,
\end{equation}
which yields
\begin{equation}
j(x_0) = \frac{1}{2 m i} 
\left[\Psi(x_0)^* \p_x \Psi(x_0) - \p_x \Psi(x_0)^* \Psi(x_0)\right] =
\frac{1}{2 m i} \Psi(x_0)^* \Psi(x_0) (\gamma^* - \gamma).
\end{equation}
This indeed vanishes, provided that $\gamma \in \R$. 

In order to investigate whether the Hamiltonian, endowed with the Robin boundary
condition eq.(\ref{bc}), is indeed self-adjoint, let us consider
\begin{eqnarray}
\langle\chi|2 m H|\Psi\rangle&=&
\int_{-\infty}^{x_0} dx \ \chi(x)^* [- \p_x^2 + 2 m V(x)] \Psi(x) 
\nonumber \\
&=&\int_{-\infty}^{x_0} dx \ [\p_x \chi(x)^* \p_x \Psi(x) + 
\chi(x)^* 2 m V(x) \Psi(x)] - \left[\chi(x)^* \p_x \Psi(x)\right]_{-\infty}^{x_0} 
\nonumber \\
&=&\int_{-\infty}^{x_0} dx \ \Psi(x) [- \p_x^2 + 2 m V(x)] \chi(x)^* \nonumber \\ 
&+&\left[\p_x \chi(x)^* \Psi(x) - \chi(x)^* \p_x \Psi(x)\right]_{-\infty}^{x_0} 
\nonumber \\
&=&\langle\Psi|2 m H|\chi\rangle^* + 
\left[\p_x \chi(x)^* \Psi(x) - \chi(x)^* \p_x \Psi(x)\right]_{-\infty}^{x_0}.
\end{eqnarray}
The Hamiltonian is Hermitean (or symmetric in mathematical parlance) if
\begin{equation}
\langle\chi|H|\Psi\rangle = \langle H^\dagger \chi|\Psi\rangle =
\langle H \chi|\Psi\rangle = \langle\Psi|H|\chi\rangle^*.
\end{equation}
This is the case, only if
\begin{equation}
\left[\p_x \chi(x)^* \Psi(x) - \chi(x)^* \p_x \Psi(x)\right]_{-\infty}^{x_0} = 0.
\end{equation}
Assuming that the wave functions are normalizable and thus vanish at infinity,
we thus conclude that Hermiticity of $H$ requires
\begin{equation}
\label{symH}
\p_x \chi(x_0)^* \Psi(x_0) - \chi(x_0)^* \p_x \Psi(x_0) = 0.
\end{equation}
Here we have also assumed that the potential $V(x)$ is not singular, such that 
all subtleties related to Hermiticity and self-adjointness are associated 
entirely with the behavior at the point $x_0$. The domain $D(H)$ of the 
Hamiltonian contains the at least twice-differentiable square-integrable wave 
functions $\Psi(x)$ that obey the Robin boundary condition eq.(\ref{bc}). Using 
that condition, eq.(\ref{symH}) reduces to
\begin{equation}
\Psi(x_0) \left[\p_x \chi(x_0)^* + \gamma \chi(x_0)^*\right] = 0.
\end{equation}
Since $\Psi(x_0)$ need not vanish, the Hamiltonian is Hermitean if
\begin{equation}
\gamma^* \chi(x_0) + \p_x \chi(x_0) = 0.
\end{equation}
Because $\gamma \in \R$, the wave function $\chi(x)$ must also obey the Robin
boundary condition eq.(\ref{bc}). Imposing this boundary condition on $\chi(x)$ 
implies that the domain of $H^\dagger$ coincides with the domain of $H$, 
$D(H^\dagger) = D(H)$. Since $H$ is indeed Hermitean when both $\Psi(x)$ and
$\chi(x)$ obey eq.(\ref{bc}), and since, in addition, $D(H^\dagger) = D(H)$, the
Hamiltonian is, in fact, self-adjoint. The parameter $\gamma$ thus characterizes
a 1-parameter family of self-adjoint extensions of the Hamiltonian on the 
half-line. 

Let us now consider the boundary condition of the corresponding supersymmetric 
partner Hamiltonian $H'$. The Robin boundary condition eq.(\ref{bc}) determines
the value of the superpotential at the boundary
\begin{equation}
W(x_0) = - \frac{\p_x \Psi(x_0)}{\Psi(x_0)} = \gamma.
\end{equation}
The eigenfunctions of the supersymmetric partner Hamiltonian $H'$ then satisfy
\begin{equation}
\Psi_n'(x_0) = \frac{1}{\sqrt{\lambda_n}} [\p_x + W(x_0)] \Psi_n(x_0) =
\frac{1}{\sqrt{\lambda_n}} [\p_x \Psi_n(x_0) + \gamma \Psi_n(x_0)] = 0,
\end{equation}
i.e.\ they obey the standard Dirichlet boundary condition, $\Psi_n'(x_0) = 0$,
which corresponds to the self-adjoint extension parameter $\gamma' = \infty$. 
In this way, non-trivial information encoded in the boundary parameter $\gamma$ 
of the original problem, is encoded in the value of the superpotential in the 
supersymmetric partner problem. In particular, all supersymmetric descendants
automatically obey the standard Dirichlet boundary condition.

\subsection{Particle on a Punctured Line}

Let us now consider a particle on a punctured line $\R \setminus \{x_0\}$, 
from which the single point $x_0$ has been removed. This divides the $x$-axis
into two separate regions $\I$ with $x > x_0$ and $\II$ with $x < x_0$. 
Accordingly, we denote the wave function and the superpotential on the two
sides of the puncture by
\begin{equation}
\Psi_{\I} = \Psi(x_0 + \varepsilon), \quad \Psi_{\II} = \Psi(x_0 - \varepsilon), 
\quad W_{\I} = W(x_0 + \varepsilon), \quad W_{\II} = W(x_0 - \varepsilon), 
\quad
\varepsilon \searrow 0.
\end{equation}
The most general boundary condition that leads to a self-adjoint Hamiltonian is
then given by
\begin{equation}
\label{bcpunct}
\left(\begin{array}{c} \Psi_{\I} \\ \p_x \Psi_{\I} \end{array} \right) = 
\exp(i \theta) \left(\begin{array}{cc} a & b \\ c & d \end{array} \right) 
\left(\begin{array}{c} \Psi_{\II} \\ \p_x \Psi_{\II} \end{array} \right), \
a, b, c, d \in \R, \ ad - bc = 1, \ \theta \in ]-\frac{\pi}{2},\frac{\pi}{2}].  
\end{equation}
The five parameters $a, b, c, d, \theta$, with the constraint $ad - bc = 1$,
define a 4-parameter family of self-adjoint extensions. The boundary condition 
(\ref{bcpunct}) guarantees that the probability current is continuous at 
$x = x_0$, i.e.
\begin{eqnarray}
2 m i \ j_{\I}\!\!&=&\!\!\Psi_{\I}^* \p_x \Psi_{\I} - \p_x \Psi_{\I}^* \Psi_{\I} =
(\Psi_{\I}^*,\p_x \Psi_{\I}^*) 
\left(\begin{array}{cc} 0 & 1 \\ -1 & 0 \end{array} \right)
\left(\begin{array}{c} \Psi_{\I} \\ \p_x \Psi_{\I} \end{array} \right) 
\nonumber \\
&=&\!\!(\Psi_{\II}^*,\p_x \Psi_{\II}^*) 
\exp(-i \theta) \left(\begin{array}{cc} a & c \\ b & d \end{array} \right)
\left(\begin{array}{cc} 0 & 1 \\ -1 & 0 \end{array} \right)
\exp(i \theta) \left(\begin{array}{cc} a & b \\ c & d \end{array} \right)
\left(\begin{array}{c} \Psi_{\II} \\ \p_x \Psi_{\II} \end{array} \right) 
\nonumber \\
&=&\!\!(\Psi_{\II}^*,\p_x \Psi_{\II}^*) 
\left(\begin{array}{cc} 0 & 1 \\ -1 & 0 \end{array} \right)
\left(\begin{array}{c} \Psi_{\II} \\ \p_x \Psi_{\II} \end{array} \right) = 
2 m i \ j_{\II}.
\end{eqnarray}
In this case, the Hermiticity of the Hamiltonian requires
\begin{equation}
\label{symHpunct}
\p_x \chi_{\I}^* \Psi_{\I} - \chi_{\I}^* \p_x \Psi_{\I} = 
\p_x \chi_{\II}^* \Psi_{\II} - \chi_{\II}^* \p_x \Psi_{\II},
\end{equation}
which implies
\begin{eqnarray}
&&\Psi_{\II}(\p_x \chi_{\I}^* \exp(i \theta) a - \chi_{\I}^* \exp(i \theta) c
- \p_x \chi_{\II}^*) + \nonumber \\
&&\p_x \Psi_{\II}(\p_x \chi_{\I}^* \exp(i \theta) b - 
\chi_{\I}^* \exp(i \theta) d + \chi_{\II}^*) = 0.
\end{eqnarray}
Since both $\Psi_{\II}$ and $\p_x \Psi_{\II}$ can take arbitrary values, we
obtain
\begin{eqnarray}
&&\chi_{\II}^* = \chi_{\I}^* \exp(i \theta) d - \p_x \chi_{\I}^* \exp(i \theta) b,
\nonumber \\
&&\p_x \chi_{\II}^* = - \chi_{\I}^* \exp(i \theta) c + 
\p_x \chi_{\I}^* \exp(i \theta) a \ \Rightarrow \nonumber \\
&&\left(\begin{array}{c} \chi_{\II} \\ \p_x \chi_{\II} \end{array} \right) = 
\exp(- i \theta) \left(\begin{array}{cc} d & -b \\ -c & a \end{array} \right) 
\left(\begin{array}{c} \chi_{\I} \\ \p_x \chi_{\I} \end{array} \right) \
\Rightarrow \nonumber \\
&&\left(\begin{array}{c} \chi_{\I} \\ \p_x \chi_{\I} \end{array} \right) = 
\exp(i \theta) \left(\begin{array}{cc} a & b \\ c & d \end{array} \right) 
\left(\begin{array}{c} \chi_{\II} \\ \p_x \chi_{\II} \end{array} \right).
\end{eqnarray}
Hence, in order to ensure Hermiticity, $\chi(x)$ must obey the same boundary
condition (\ref{bcpunct}) as $\Psi(x)$. This ensures that the domain of
$H^\dagger$ coincides with the domain of $H$, $D(H^\dagger) = D(H)$, which implies
that $H$ is not only Hermitean, but actually self-adjoint.

\subsection{Parity Symmetry}

Let us investigate under what circumstances the most general self-adjoint point 
interaction at $x_0 = 0$ respects the parity symmetry, generated by
$^P \Psi(x) = \Psi(-x)$, which implies
\begin{equation}
^P \Psi_{\I} = \Psi_{\II}, \quad ^P \Psi_{\II} = \Psi_{\I}, \quad
^P \p_x \Psi_{\I} = - \p_x \Psi_{\II}, \quad ^P \p_x \Psi_{\II} = - \p_x\Psi_{\I}.
\end{equation}
The point interaction is parity-symmetric if the parity partner $^P \Psi(x)$ of 
a wave function $\Psi(x)$ also respects the boundary condition, i.e.
\begin{eqnarray}
&&\left(\begin{array}{c} ^P \Psi_{\I} \\ ^P \p_x \Psi_{\I} \end{array} \right) = 
\exp(i \theta) \left(\begin{array}{cc} a & b \\ c & d \end{array} \right) 
\left(\begin{array}{c} ^P \Psi_{\II} \\ ^P \p_x \Psi_{\II} \end{array} \right) \
\Rightarrow \nonumber \\
&&\left(\begin{array}{c} \Psi_{\II} \\ - \p_x \Psi_{\II} \end{array} \right) = 
\exp(i \theta) \left(\begin{array}{cc} a & b \\ c & d \end{array} \right) 
\left(\begin{array}{c} \Psi_{\I} \\ - \p_x \Psi_{\I} \end{array} \right) \
\Rightarrow \nonumber \\
&&\left(\begin{array}{cc} 1 & 0 \\ 0 & -1 \end{array} \right) 
\left(\begin{array}{c} \Psi_{\II} \\ \p_x \Psi_{\II} \end{array} \right) = 
\exp(i \theta) \left(\begin{array}{cc} a & b \\ c & d \end{array} \right) 
\left(\begin{array}{cc} 1 & 0 \\ 0 & -1 \end{array} \right) 
\left(\begin{array}{c} \Psi_{\I} \\ \p_x \Psi_{\I} \end{array} \right) \
\Rightarrow \nonumber \\
&&\left(\begin{array}{c} \Psi_{\I} \\ \p_x \Psi_{\I} \end{array} \right) =
\exp(- i \theta) \left(\begin{array}{cc} 1 & 0 \\ 0 & -1 \end{array} \right) 
\left(\begin{array}{cc} d & -b \\ -c & a \end{array} \right) 
\left(\begin{array}{cc} 1 & 0 \\ 0 & -1 \end{array} \right) 
\left(\begin{array}{c} \Psi_{\II} \\ \p_x \Psi_{\II} \end{array} \right) \
\Rightarrow \ \nonumber \\
&&\left(\begin{array}{c} \Psi_{\I} \\ \p_x \Psi_{\I} \end{array} \right) =
\exp(- i \theta) \left(\begin{array}{cc} d & b \\ c & a \end{array} \right) 
\left(\begin{array}{c} \Psi_{\II} \\ \p_x \Psi_{\II} \end{array} \right).
\end{eqnarray}
This is consistent with the original boundary condition (\ref{bcpunct}) only
if $a = d$ and $\theta = 0$.

\subsection{Self-Adjointness of the Superpartner Hamiltonian}

Let us now investigate the superpartner Hamiltonian $H'$ for a particle on the 
punctured line, without assuming parity symmetry. First, we notice that
(for $n > 0$)
\begin{equation}
\Psi_n(x) = \frac{1}{\lambda_n} A^\dagger A \Psi_n(x) = 
\frac{1}{\sqrt{\lambda_n}} A^\dagger \Psi_n'(x) = 
\frac{1}{\sqrt{\lambda_n}}[- \p_x \Psi_n'(x) + W(x) \Psi_n'(x)],
\end{equation}
such that
\begin{eqnarray}
\p_x \Psi_n'(x)&=&W(x) \Psi_n'(x) - \sqrt{\lambda_n} \Psi_n(x) \nonumber \\
&=&W(x) \frac{1}{\sqrt{\lambda_n}} [\p_x \Psi_n(x) + W(x) \Psi_n(x)] - 
\sqrt{\lambda_n} \Psi_n(x) \nonumber \\
&=&\frac{1}{\sqrt{\lambda_n}} [W(x) \p_x \Psi_n(x) + (W(x)^2 - \lambda_n) 
\Psi_n(x)],
\end{eqnarray}
which implies
\begin{equation}
\label{susyrelation}
\left(\begin{array}{c} \Psi_n'(x) \\ \p_x \Psi_n'(x) \end{array} \right) =
\frac{1}{\sqrt{\lambda_n}} \left(\begin{array}{cc} W(x) & 1 \\ 
W(x)^2 - \lambda_n & W(x) \end{array} \right) 
\left(\begin{array}{c} \Psi_n(x) \\ \p_x \Psi_n(x) \end{array} \right).
\end{equation}

Based on the previous discussion, for the superpartner we again expect a 
boundary condition of the form
\begin{equation}
\left(\begin{array}{c} \Psi'_{\I} \\ \p_x \Psi'_{\I} \end{array} \right) = 
\exp(i \theta') \left(\begin{array}{cc} a' & b' \\ c' & d' \end{array} \right) 
\left(\begin{array}{c} \Psi'_{\II} \\ \p_x \Psi'_{\II} \end{array} \right), \
a', b', c', d' \in \R, \ a'd' - b'c' = 1.
\end{equation}
Using eq.(\ref{susyrelation}), which immediately implies 
\begin{eqnarray}
&&\left(\begin{array}{c} \Psi_{n\I}' \\ \p_x \Psi_{n\I}' \end{array} \right) =
\frac{1}{\sqrt{\lambda_n}} \left(\begin{array}{cc} W_{\I} & 1 \\ 
W_{\I}^2 - \lambda_n & W_{\I} \end{array} \right) 
\left(\begin{array}{c} \Psi_{n\I} \\ \p_x \Psi_{n\I} \end{array} \right), 
\nonumber \\
&&\left(\begin{array}{c} \Psi_{n\II}' \\ \p_x \Psi_{n\II}' \end{array} \right) =
\frac{1}{\sqrt{\lambda_n}} \left(\begin{array}{cc} W_{\II} & 1 \\ 
W_{\II}^2 - \lambda_n & W_{\II} \end{array} \right) 
\left(\begin{array}{c} \Psi_{n\II} \\ \p_x \Psi_{n\II} \end{array} \right),
\end{eqnarray}
as well as eq.(\ref{bcpunct}), one identifies
\begin{equation}
\label{matrix}
\exp(i \theta') \left(\begin{array}{cc} a' & b' \\ c' & d' \end{array} \right)
= \left(\begin{array}{cc} W_{\I} & 1 \\ W_{\I}^2 - \lambda_n & W_{\I} \end{array} 
\right) 
\exp(i \theta) \left(\begin{array}{cc} a & b \\ c & d \end{array} \right)
\left(\begin{array}{cc} W_{\II} & 1 \\ W_{\II}^2 - \lambda_n & W_{\II} \end{array} 
\right)^{-1}.
\end{equation}
It is important to note that the values of the superpotential $W_{\I}$ and
$W_{\II}$ at the two sides of the puncture $x_0$ are not independent, but are
related by
\begin{eqnarray}
\Psi_{0\I} \left(\begin{array}{c} 1 \\ - W_{\I} \end{array}\right)&=&
\left(\begin{array}{c} \Psi_{0\I} \\ \p_x \Psi_{0\I} \end{array}\right) =
\exp(i \theta) \left(\begin{array}{cc} a & b \\ c & d \end{array} \right)
\left(\begin{array}{c} \Psi_{0\II} \\ \p_x \Psi_{0\II} \end{array}\right) 
\nonumber \\
&=&\exp(i \theta) \left(\begin{array}{cc} a & b \\ c & d \end{array} \right)
\Psi_{0\II} \left(\begin{array}{c} 1 \\ - W_{\II} \end{array}\right) =
\exp(i \theta) \Psi_{0\II} 
\left(\begin{array}{c} a - b W_{\II} \\ c - d W_{\II} \end{array}\right), 
\nonumber \\ \,
\end{eqnarray}
such that
\begin{equation}
\label{constraint}
\frac{1}{- W_{\I}} = \frac{a - b W_{\II}}{c - d W_{\II}} \ \Rightarrow \
a W_{\I} - b W_{\I} W_{\II} + c - d W_{\II} = 0,
\end{equation}
Using this relation, it is straightforward to work out the individual matrix 
elements in eq.(\ref{matrix}) and one obtains
\begin{eqnarray}
\label{susybc}
&&\exp(i \theta') a' = \exp(i \theta) (d + b W_{\I}), \nonumber \\
&&\exp(i \theta') b' = 0, \nonumber \\
&&\exp(i \theta') c' = \exp(i \theta)
[(d + b W_{\I}) W_{\I} - (a - b W_{\II}) W_{\II} - b \lambda_n], \nonumber \\
&&\exp(i \theta') d' = \exp(i \theta) (a - b W_{\II}).
\end{eqnarray}
Let us check the constraint
\begin{equation}
a'd' - b'c' = (d + b W_{\I})(a - b W_{\II}) = 
a d + a b W_{\I} - b d W_{\II} - b^2 W_{\I} W_{\II} = a d - b c = 1,
\end{equation}
which is indeed correctly satisfied.
Since, the parameters $a', b', c', d', \theta'$ must be the same for every 
state, it is unacceptable that the eigenvalue $\lambda_n$ enters the expression 
for $c'$ in eq.(\ref{susybc}). In fact, the $\lambda_n$-dependence of the
boundary condition implies that the superpartner Hamiltonian $H'$ is not 
self-adjoint, unless $b = 0$. This means that only a 3-parameter sub-family of
self-adjoint Hamiltonians $H$ (namely those with $b = 0$) have a self-adjoint
superpartner $H'$. In that case, the constraint $a d - b c = a d = 1$ implies
$d = 1/a$. The self-adjoint extension parameters of the supersymmetric 
descendant then are
\begin{equation}
\label{bcpartner}
a' = d = 1/a, \quad b' = 0, \quad c' = d W_{\I} - a W_{\II} = W_{\I}/a - a W_{\II},
\quad d' = a, \quad \theta' = \theta.
\end{equation}
For $b = 0$ eq.(\ref{constraint}) reduces to $c = W_{\II}/a - a W_{\I}$.
It is interesting to note that $b = 0$ implies that the probability density
$\rho(x) = |\Psi(x)|^2$ (but not necessarily the wave function itself) is
continuous at the puncture $x_0$. This is thus a necessary and sufficient 
condition for the self-adjointness of the superpartner. Since $b' = 0$ for the 
superpartner, all higher supersymmetric descendants are then also automatically 
self-adjoint. When the original boundary condition is parity-invariant, i.e.\ 
when $a = d = 1/a = \pm 1$ and $\theta = 0$, the superpartner also obeys a 
parity-symmetric boundary condition with
\begin{equation}
a' = a = \pm 1, \quad b' = 0, \quad c' = W_{\I}/a - a W_{\II} = - c,
\quad d' = a, \quad \theta' = \theta = 0.
\end{equation}
Here we have used 
$a W_{\I} + b W_{\I} W_{\II} + c - d W_{\II} = W_{\I}/a - a W_{\II} + c = 0$.

\subsection{Self-Adjointness of the Operator ${\cal Q}_+$}

Let us now investigate the self-adjoint extensions of the operator ${\cal Q}_+$.
Introducing the 2-component wave functions
\begin{equation}
\widetilde{\chi}(x) = 
\left(\begin{array}{c} \chi(x) \\ \chi'(x) \end{array} \right), \quad
\widetilde{\Psi}(x) = 
\left(\begin{array}{c} \Psi(x) \\ \Psi'(x) \end{array} \right),
\end{equation}
for the particle on a half-line we obtain
\begin{eqnarray}
\langle \widetilde{\chi}|{\cal Q}_+|\widetilde{\Psi} \rangle&=&
\int_{-\infty}^{x_0} dx \ (\chi(x)^*,\chi'(x)^*) 
\left(\begin{array}{cc} 0 & - \p_x + W(x) \\ \p_x + W(x) & 0 \end{array} \right)
\left(\begin{array}{c} \Psi(x) \\ \Psi'(x) \end{array} \right) \nonumber \\
&=&\int_{-\infty}^{x_0} dx \ 
\{\chi(x)^*[- \p_x + W(x)] \Psi'(x) + \chi'(x)^* [\p_x + W(x)] \Psi(x)\}
\nonumber \\
&=&\int_{-\infty}^{x_0} dx \
\{\p_x \chi(x)^* \Psi'(x) + \chi(x)^* W(x) \Psi'(x) \nonumber \\
&-&\p_x \chi'(x)^* \Psi(x) + \chi'(x)^* W(x) \Psi(x)\} -
[\chi(x)^* \Psi'(x) - \chi'(x)^* \Psi(x)]_{-\infty}^{x_0} \nonumber \\
&=&\langle \widetilde{\Psi}|{\cal Q}_+|\widetilde{\chi} \rangle^* -
\chi(x_0)^* \Psi'(x_0) + \chi'(x_0)^* \Psi(x_0).
\end{eqnarray}
Hermiticity of ${\cal Q}_+$ thus requires 
$\chi(x_0)^* \Psi'(x_0) = \chi'(x_0)^* \Psi(x_0)$. We make the ansatz
\begin{equation}
\Psi'(x_0) = \eta \Psi(x_0),
\end{equation}
for the self-adjoint extension condition, such that Hermiticity of ${\cal Q}_+$
then requires
\begin{equation}
\chi(x_0)^* \eta \Psi(x_0) = \chi'(x_0)^* \Psi(x_0) \ \Rightarrow \
\chi'(x_0) = \eta^* \chi(x_0).
\end{equation}
Hence, the domains of ${\cal Q}_+$ and its adjoint ${\cal Q}_+^\dagger$ agree, 
i.e.\ $D({\cal Q}_+) = D({\cal Q}_+^\dagger)$, if and only if $\eta = \eta^*$, 
which ensures that ${\cal Q}_+$ is not only Hermitean but actually 
self-adjoint. There is a 1-parameter family of self-adjoint extensions of 
${\cal Q}_+$ (parameterized by $\eta \in \R$).

From our previous considerations we know that 
$\p_x \Psi(x_0) + \gamma \Psi(x_0) = 0$ implies $\Psi'(x_0) = 0$, which in turn 
leads to $\eta = 0$. This seems to leave the value $\Psi(x_0)$ unrestricted.
It also seems that ${\cal Q}_+$ contains no free parameter, while $H$ is endowed
with the Robin boundary condition characterized by $\gamma$. This apparent
contradiction gets resolved when we recall that $\gamma$ is also encoded in the
superpotential, i.e.\ $W(x_0) = \gamma$. Hence, we may conclude that 
${\cal Q}_+$ is indeed self-adjoint, but has a fixed extension parameter 
$\eta = 0$. Why are we not encountering the other self-adjoint extensions of 
${\cal Q}_+$? Actually, in our treatment of $H$ and $H'$ those are unphysical. 
We should point out that we consider $H$ and its supersymmetric descendant $H'$ 
as two physically distinct systems, rather than as two parts of the same system.
In particular, we do not allow states with both an upper and a lower component.
Consequently, probability is conserved separately for the systems associated 
with $H$ and $H'$. This limits us to the self-adjoint extension of ${\cal Q}_+$
characterized by $\eta = 0$.

Let us also consider the self-adjoint extensions of the operator ${\cal Q}_+$ 
on the punctured line, consisting of the regions $\I$ and $\II$. In that case, 
Hermiticity requires
\begin{equation}
\chi_{\I}^* \Psi'_{\I} - {\chi'_{\I}}^* \Psi_{\I} =
\chi_{\II}^* \Psi'_{\II} - {\chi'_{\II}}^* \Psi_{\II}.
\end{equation}
We now make the ansatz
\begin{equation}
\left(\begin{array}{c} \Psi_{\I} \\ \Psi'_{\I} \end{array}\right) =
\left(\begin{array}{cc} \mu & \nu \\ \rho & \sigma \end{array}\right)
\left(\begin{array}{c} \Psi_{\II} \\ \Psi'_{\II} \end{array}\right),
\end{equation}
for the self-adjoint extension condition, such that Hermiticity then requires
\begin{eqnarray}
&&\chi_{\I}^* \Psi'_{\I} - {\chi'_{\I}}^* \Psi_{\I} =
\chi_{\I}^* (\rho \Psi_{\II} + \sigma \Psi'_{\II}) - 
{\chi'_{\I}}^* (\mu \Psi_{\II} + \nu \Psi'_{\II}) =
\chi_{\II}^* \Psi'_{\II} - {\chi'_{\II}}^* \Psi_{\II} \ \Rightarrow \nonumber \\
&&\Psi_{\II} (\rho \chi_{\I}^* - \mu {\chi'_{\I}}^* + {\chi'_{\II}}^*) +
\Psi'_{\II} (\sigma \chi_{\I}^* - \nu {\chi'_{\I}}^* - \chi_{\II}^*) = 0 \ 
\Rightarrow \nonumber \\
&&\left(\begin{array}{c} \chi_{\II} \\ \chi'_{\II} \end{array}\right) =
\left(\begin{array}{cc} \sigma^* & -\nu^* \\ -\rho^* & \mu^* \end{array}\right)
\left(\begin{array}{c} \chi_{\I} \\ \chi'_{\I} \end{array}\right) \
\Rightarrow \nonumber \\
&&\left(\begin{array}{c} \chi_{\I} \\ \chi'_{\I} \end{array}\right) =
\frac{1}{\mu^* \sigma^* - \nu^* \rho^*}
\left(\begin{array}{cc} \mu^* & \nu^* \\ \rho^* & \sigma^* \end{array}\right)
\left(\begin{array}{c} \chi_{\II} \\ \chi'_{\II} \end{array}\right).
\end{eqnarray}
The operator ${\cal Q}_+$ is self-adjoint if it is Hermitean and if 
$D({\cal Q}_+) = D({\cal Q}_+^\dagger)$, which is the case if $\mu$, $\nu$, 
$\rho$, and $\sigma$ are real, up to a common complex phase, and if 
$|\mu \sigma - \nu \rho| = 1$. Hence, there is a 4-parameter family of 
self-adjoint extensions of ${\cal Q}_+$.

From our previous investigations we know that the supersymmetric descendant $H'$
is self-adjoint only if the self-adjoint extension of the original Hamiltonian
$H$ obeys $b = 0$, which implies
\begin{equation}
\Psi_{\I} = \exp(i \theta) a \Psi_{\II}, \quad  
\Psi'_{\I} = \exp(i \theta) \frac{1}{a} \Psi'_{\II}.
\end{equation}
This in turn leads to
\begin{equation}
\mu = \exp(i \theta) a, \ \nu = \rho = 0, \ \sigma = \exp(i \theta) \frac{1}{a},
\end{equation}
which means that the operator ${\cal Q}_+$ is indeed self-adjoint. However, as 
in the case of the half-line, we are not encountering all possible self-adjoint
extensions of ${\cal Q}_+$. In fact, we are only exploring a 2-parameter 
sub-class (parametrized by $\mu$ and $\sigma$ or equivalently $a$ and $\theta$)
of the general 4-parameter family of self-adjoint extensions of ${\cal Q}_+$. 
Again, the other self-adjoint extensions are unphysical in our context, because 
$H$ and $H'$ describe distinct physical systems. Again, one may ask how a 
2-parameter family of self-adjoint extensions of ${\cal Q}_+$ can lead to the 
3-parameter family associated with $H$ and $H'$ or equivalently 
${\cal Q}_+^2 = {\cal H}$. Again, this is because the information about the 
third self-adjoint extension parameter (in this case $c$) is encoded in the 
superpotential as $c = W_{\II}/a - a W_{\I}$, or equivalently 
$c' = W_{\I}/a - a W_{\II}$.

Our investigation of the self-adjointness of ${\cal Q}_+$ also sheds more light
on the question why we had to put $b = 0$. While we encountered this condition
when we demanded the self-adjointness of $H'$, it would also have followed from
requiring self-adjointness of ${\cal Q}_+$. Indeed, for $b \neq 0$ the 
self-adjoint extension condition for $H$ leads to
\begin{equation}
\Psi_{\I} = \exp(i \theta)(a \Psi_{\II} + b \p_x \Psi_{\II}). 
\end{equation}
This condition is inconsistent with the most general self-adjoint extension of
${\cal Q}_+$, which implies
\begin{equation}
\Psi_{\I} = \mu \Psi_{\II} + \nu \Psi'_{\II}.
\end{equation}
The two conditions are consistent only for $\mu = \exp(i \theta) a$ and 
$b = \nu = 0$.

\section{Applications to Concrete Problems}

In this section, we illustrate the general results of the previous section by
considering several concrete examples: an otherwise free particle, subject to a 
point interaction, a particle in a box, and a combination of these two cases. 
The latter displays a bulk-boundary resonance, whose manifestation in the 
supersymmetric descendant is also investigated.

\subsection{Point Interaction with Non-Self-Adjoint Superpartner}

Let us consider an otherwise free particle moving on the punctured line 
$\R \setminus \{0\}$, subject to a parity-invariant point interaction,
characterized by the self-adjoint extension parameters $a, b, c, d = a, 
\theta = 0$. In this subsection, we allow $b \neq 0$, which, based on the 
results of the previous section, is expected not to lead to a self-adjoint
superpartner $H'$. A bound state wave function for the even (+) and odd (-)
parity states can be written as
\begin{equation}
\Psi_\pm(x) = C_\pm \exp(- \varkappa_\pm x), \ x > 0, \quad 
\Psi_\pm(x) = \pm C_\pm \exp(\varkappa_\pm x), \ x < 0, \quad 
\varkappa_\pm \geq 0.
\end{equation}
The boundary condition (\ref{bcpunct}) then implies
\begin{equation}
\left(\begin{array}{c} 1 \\ - \varkappa_\pm \end{array} \right) = 
\pm \left(\begin{array}{cc} a & b \\ c & a \end{array} \right)
\left(\begin{array}{c} 1 \\ \varkappa_\pm \end{array} \right) = 
\pm \left(\begin{array}{c} a + b \varkappa_\pm \\ c + a \varkappa_\pm \end{array}
\right),
\end{equation}
such that (for $b \neq 0$)
\begin{equation}
\frac{1}{- \varkappa_\pm} = \frac{a + b \varkappa_\pm}{c + a \varkappa_\pm} \ 
\Rightarrow \
\varkappa_{\pm} = \frac{1}{b}(- a \pm 1).
\end{equation}
Here we have used $a d - b c = a^2 - b c = 1$. Let us assume that $b > 0$. Then
for $a < - 1$ there are two bound states, for $-1 \leq a < 1$ there is one
bound state, and for $a \geq 1$ there is no bound state. We consider the
case with two bound states, i.e.\ $a < - 1$, $b > 0$. Then the ground state is
even under parity with the wave function
\begin{equation}
\Psi_0(x) = \Psi_+(x) = C_+ \exp(- \varkappa_+ |x|),
\end{equation}
and the first excited state (which is also bound) is parity-odd with the 
wave function
\begin{equation}
\Psi_1(x) = \Psi_-(x) = C_- \sign(x) \exp(- \varkappa_-|x|).
\end{equation}
The corresponding eigenvalues of $H$ are given by
\begin{equation}
\lambda_0 = 0, \quad \lambda_1 = \varkappa_+^2 - \varkappa_-^2 = 
- \frac{4 a}{b^2}.
\end{equation}

The ground state wave function gives rise to the superpotential
\begin{equation}
W(x) = - \p_x \log \Psi_0(x) = \sign(x) \varkappa_+,
\end{equation}
and thus to the superpartner potential
\begin{equation}
2 m V'(x) = \p_x W(x) + W(x)^2 = \varkappa_+^2.
\end{equation}
Note that the original potential is given by the same constant
\begin{equation}
2 m V(x) = - \p_x W(x) + W(x)^2 = \varkappa_+^2.
\end{equation}
This is consistent with the shift that ensures a vanishing ground state energy,
i.e.\ $\lambda_0 = 0$. 

Let us now construct the ground state of the superpartner Hamiltonian $H'$, 
which we expect not to be self-adjoint because $b \neq 0$,
\begin{equation}
\Psi_1'(x) = \frac{1}{\sqrt{\lambda_1}} [\p_x + W(x)] \Psi_1(x) = 
\frac{C_-}{\sqrt{- a}} \exp(- \varkappa_- |x|).
\end{equation}
Unless $a = - 1$, this wave function is, in fact, incorrectly normalized, which
already indicates that, for $b \neq 0$, something is wrong with the construction
of $H'$. 

Let us also consider the scattering states of the original Hamiltonian. For 
simplicity, we consider parity eigenstates, although this does not correspond
to the standard scattering geometry with an incident wave coming from only one
direction. An ansatz for a parity-odd scattering state is given by
\begin{equation}
\Psi_{k-}(x) = D_- \sign(x) \cos(k|x| + \delta), \quad
\p_x \Psi_{k-}(x) = - D_- k \sin(k|x| + \delta),
\end{equation}
which yields
\begin{equation}
\left(\begin{array}{c} \cos\delta \\ - k \sin\delta \end{array} \right) = 
\left(\begin{array}{cc} a & b \\ c & a \end{array} \right)
\left(\begin{array}{c} - \cos\delta \\ - k \sin\delta \end{array} \right) \
\Rightarrow \ \tan\delta = - \frac{a + 1}{b k} = \frac{\varkappa_-}{k}.
\end{equation}
This scattering state is indeed orthogonal to the two bound states, as it must
be, because $H$ is self-adjoint
\begin{eqnarray}
\langle \Psi_1|\Psi_{k-}\rangle&=&2 C_- D_- 
\int_0^\infty dx \ \cos(k x + \delta) \exp(- \varkappa_- x) \nonumber \\
&=&\frac{\varkappa_- \cos\delta - k \sin\delta}{k^2 + \varkappa_-^2} = 
\frac{\cos\delta}{k^2 + \varkappa_-^2}(\varkappa_- - k \tan\delta) = 0.
\end{eqnarray}
Let us now construct the corresponding supersymmetric partner wave function of 
the scattering state
\begin{equation}
\Psi_{k-}'(x) = \frac{1}{\sqrt{\lambda}} [\p_x + W(x)] \Psi_{k-}(x) = 
\frac{D_-}{\sqrt{k^2 + \varkappa_+^2}} 
[\varkappa_+ \cos(k|x| + \delta) - k \sin(k|x| + \delta)].
\end{equation}
Here we have used $\lambda = k^2 + \varkappa_+^2$. This wave function is, in 
fact, not orthogonal to the corresponding bound state because
\begin{eqnarray}
\langle \Psi_1'|\Psi_{k-}'\rangle&=&\frac{2 C_- D_-}
{\sqrt{- a(k^2 + \varkappa_+^2)}} 
\int_0^\infty dx \ [\varkappa_+ \cos(k x + \delta) - k \sin(k x + \delta)]
\exp(- \varkappa_- x) \nonumber \\
&=&- \frac{2 C_- D_-}{\sqrt{- a(k^2 + \varkappa_+^2)}} \cos\delta \neq 0.
\end{eqnarray}
This confirms that $H'$ cannot be self-adjoint, which is what we expected, 
since we chose $b \neq 0$.

\subsection{General Point Interaction with a Self-Adjoint Superpartner}

Let us now put $b = 0$, in which case both $H$ and its superpartner $H'$ are
self-adjoint. In this case, $a d - b c = a d = 1$, such that
$d = 1/a$. The most general, not necessarily parity-symmetric, point interaction
is then characterized by the three parameters $a, c, \theta$. First, we look for
a bound state
\begin{equation}
\Psi_0(x) = C_{\I} \exp(- \varkappa x), \ x > 0, \quad
\Psi_0(x) = C_{\II} \exp(\varkappa x), \ x < 0,
\end{equation}
which implies
\begin{equation}
C_{\I} \left(\begin{array}{c} 1 \\ - \varkappa \end{array} \right) = 
\exp(i \theta) \left(\begin{array}{cc} a & 0 \\ c & 1/a \end{array} \right) 
C_{\II} \left(\begin{array}{c} 1 \\ \varkappa \end{array} \right) \
\Rightarrow \ \varkappa = - \frac{a c}{1 + a^2}.
\end{equation}
A bound state exists only for $a c < 0$. The corresponding superpotential is 
then given by
\begin{equation}
W(x) = - \p_x \log \Psi_0(x) = \sign(x) \varkappa \ \Rightarrow \
2 m V(x) = 2 m V'(x) = \varkappa^2.
\end{equation}
According to eq.(\ref{bcpartner}), the self-adjoint extension parameters for the
superpartner are
\begin{equation}
a' = d = \frac{1}{a}, \quad b' = 0, \quad 
c' = d W_{\I} - a W_{\II} = \varkappa \left(\frac{1}{a} + a \right) = - c, \quad 
d' = a, \quad \theta' = \theta.
\end{equation}
Since $a'c' = - c/a = a c/a^2 > 0$, if $H$ has a bound state, the superpartner 
$H'$ does not.

When we demand parity symmetry, we require $\theta = 0$, $d = 1/a = a$, such
that $a = \pm 1$. For $a = 1$ the problem corresponds to the standard 
$\delta$-function potential. The corresponding wave function, illustrated in 
the left panel of figure 1, is then continuous at the puncture $x_0 = 0$. It is 
interesting to note that, for $a = -1$, the ground state, illustrated in the 
right panel of figure 1, is parity-odd.
\begin{figure}[t]
\begin{center}
\epsfig{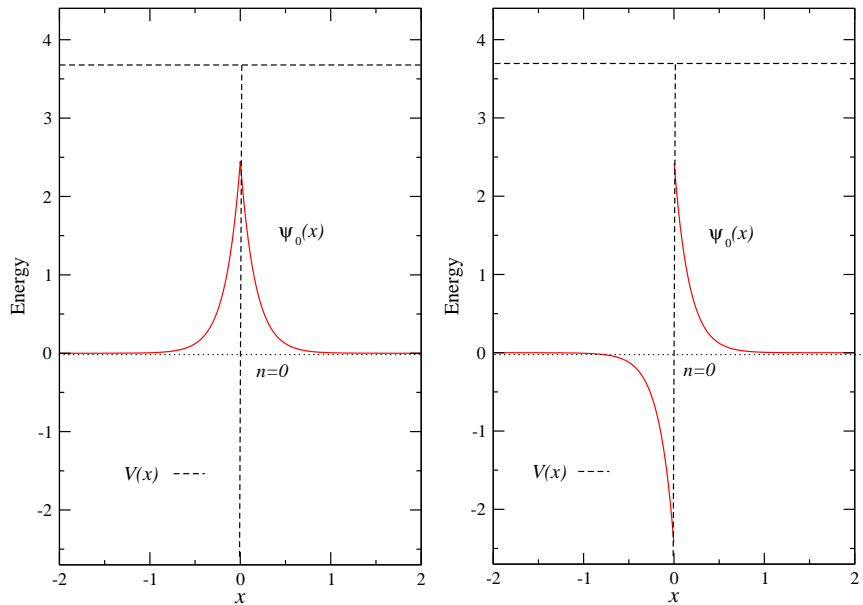}
\end{center}
\caption{\it Left panel: Ground state wave function for a parity-invariant
point interaction with $a = d = 1$, $b = 0$, $\theta = 0$. The horizontal
dashed lines indicate the potential $V(x)$. The downward vertical dashed line 
symbolizes an attractive point interaction at $x_0 = 0$, equivalent to a
$\delta$-function potential. Right panel: Ground state wave function for a 
parity-invariant point interaction with $a = d = - 1$, $b = 0$, $\theta = 0$.}
\end{figure}

Let us also consider scattering states, first for $H$, and no longer assuming
parity symmetry. We make the ansatz
\begin{equation}
\Psi_k(x) = \exp(i k x) + R \exp(- i k x), \ x < 0, \quad 
\Psi_k(x) = T \exp(i k x), \ x > 0, 
\end{equation}
which implies
\begin{eqnarray}
&&\left(\begin{array}{c} T \\ i k T \end{array} \right) = 
\exp(i \theta) \left(\begin{array}{cc} a & 0 \\ c & 1/a \end{array} \right) 
\left(\begin{array}{c} 1 + R \\ i k (1 - R) \end{array} \right) \ \Rightarrow 
\nonumber \\
&&R = \frac{i k (1 - a^2) + a c}{i k (1 + a^2) - a c}, \quad
T = \frac{2 i k a \exp(i \theta)}{i k (1 + a^2) - a c}. 
\end{eqnarray}
Similarly, for the superpartner one obtains
\begin{eqnarray}
&&R' = \frac{i k (1 - {a'}^2) + a' c'}{i k (1 + {a'}^2) - a' c'}
= \frac{i k (a^2 - 1) - a c}{i k (a^2 + 1) + a c}, \nonumber \\
&&T' = \frac{2 i k a' \exp(i \theta)}{i k (1 + {a'}^2) - a' c'} =
\frac{2 i k a \exp(i \theta)}{i k (a^2 + 1) + a c}. 
\end{eqnarray}
In particular, one obtains $|R'| = |R|$, $|T'| = |T|$.

\subsection{Self-Adjoint Extensions of a Particle in a Box}

Let us now consider a particle in a box of size $L$ ($x \in [-L/2,L/2]$) with a
general Robin boundary condition characterized by the self-adjoint extension
parameter $\gamma$. In order to ensure parity symmetry, we impose the boundary
condition in a symmetric manner such that
\begin{equation}
\p_x \Psi(L/2) + \gamma \Psi(L/2) = 0, \quad
\p_x \Psi(-L/2) - \gamma \Psi(-L/2) = 0.
\end{equation}
The Dirichlet boundary condition that is used in standard textbook treatments
corresponds to $\gamma = \infty$. A detailed discussion for general values of 
$\gamma$ can be found in \cite{AlH12}. The resulting spectra and wave functions
are illustrated in figure 2.
\begin{figure}[t]
\begin{center}
\vskip-1.5cm
\includegraphics[width=0.4\textwidth,angle=-90]{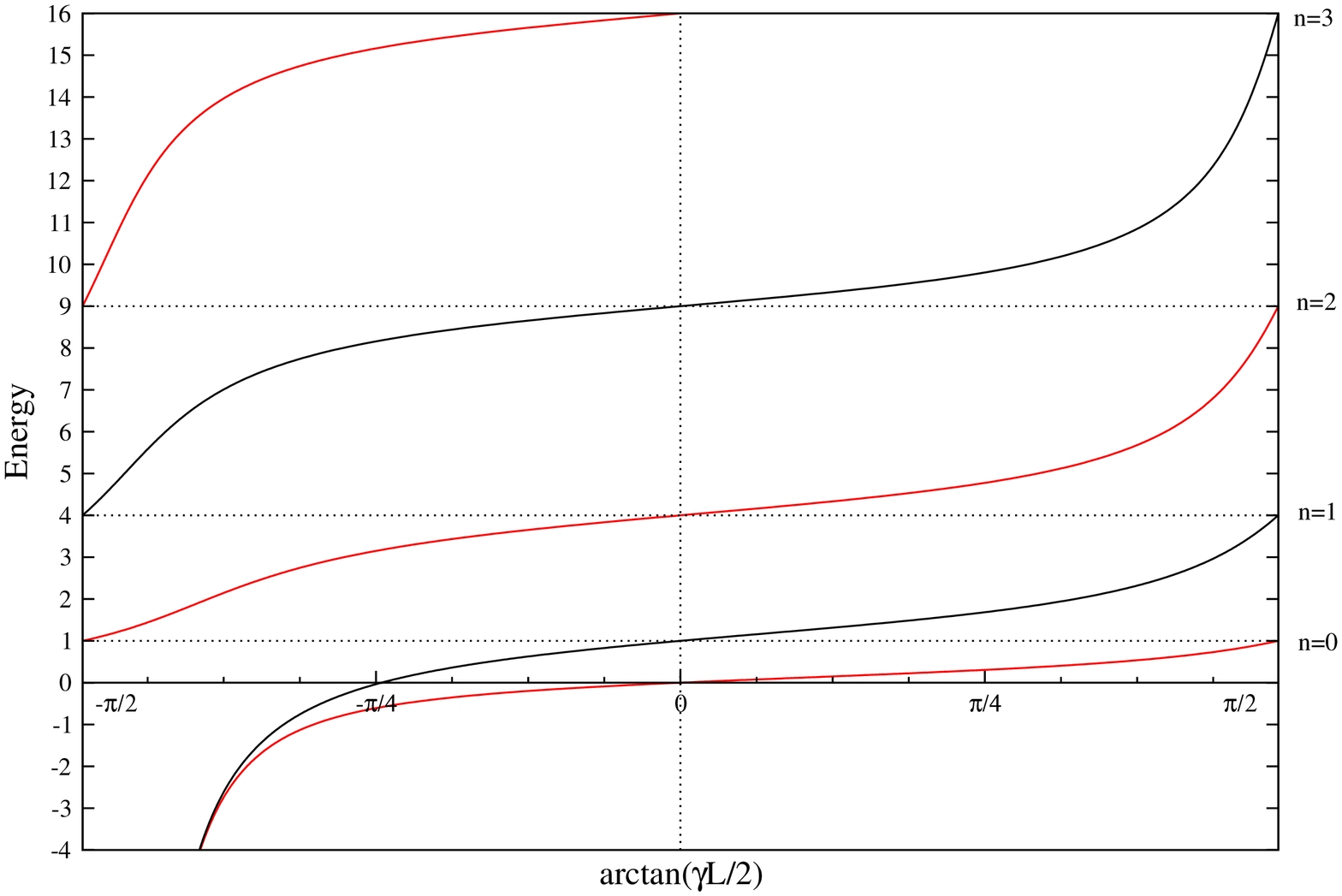} \\ \vskip0.5cm
\includegraphics[width=0.48\textwidth,angle=-90]{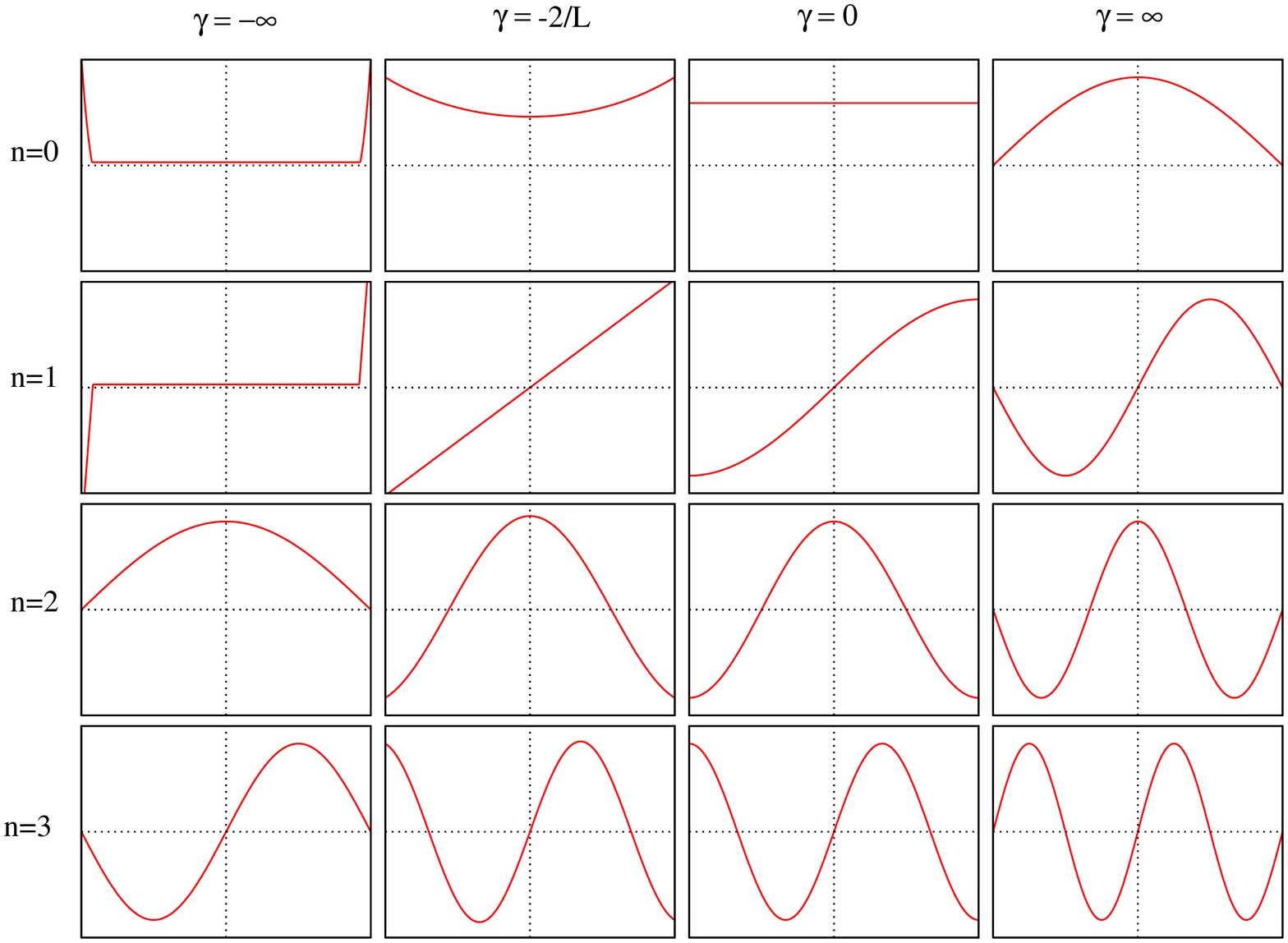}
\end{center}
\caption{\it Top panel: Energy spectrum of a particle in a box as a function
of the self-adjoint extension parameter $\gamma$. The $x$-value represents
$\arctan(\gamma L/2) \in [- \tfrac{\pi}{2},\tfrac{\pi}{2}]$, which corresponds 
to $\gamma \in [-\infty,\infty]$. The $y$-value represents the energies $E_n$ 
(with $n = 0, 1, 2, 3, 4$) in units of $\pi^2/(2 m L^2)$. Bottom panel: The 
wave functions $\Psi_n(x)$, $x \in [- \tfrac{L}{2},\tfrac{L}{2}]$, (with
$n = 0, 1, 2, 3$), for $\gamma = - \infty, - \tfrac{2}{L}, 0, \infty$. The 
sharp peaks in the $n=0$ and $n=1$ states at $\gamma = - \infty$ represent 
$\delta$-function-type wave functions of negative infinite energy localized at 
the boundaries. Except for these states, the energies and wave functions at 
$\gamma = - \infty$ are the same as those at $\gamma = \infty$.} 
\end{figure}
The case $\gamma = 0$ corresponds to Neumann boundary conditions, for 
which the ground state has zero energy. For $\gamma < 0$, there are even 
negative energy states, localized at the wall. This seems to contradict the
Heisenberg uncertainty relation, which, however, applies only to the infinite
volume. A generalized uncertainty relation, which applies to a finite volume,
was derived in \cite{AlH12}, and is consistent with negative energy values.

For $\gamma > 0$, the even parity states take the form
\begin{equation}
\Psi_n(x) = B \cos(k_n x), \ 
\frac{\gamma}{k_n} = \tan(k_n L/2), \ n = 0,2,4,\dots,
\end{equation}
while the odd parity states are given by
\begin{equation}
\label{sin}
\Psi_n(x) = C \sin(k_n x), \ 
\frac{\gamma}{k_n} = - \cot(k_n L/2), \ n = 1,3,5,\dots.
\end{equation}
The superpotential thus takes the form
\begin{equation}
W(x) = - \frac{\p_x \Psi_0(x)}{\Psi_0(x)} = k_0 \tan(k_0 x),
\end{equation}
such that
\begin{eqnarray}
&&2 m V(x) = - \p_x W(x) + W(x)^2 = - k_0^2, \nonumber \\
&&2 m V'(x) = \p_x W(x) + W(x)^2 = 
k_0^2  \left(\frac{2}{\cos^2(k_0 x)} - 1 \right).
\end{eqnarray}
The ground state of the first supersymmetric descendant is obtained as
\begin{equation}
\Psi_1'(x) = \frac{1}{\sqrt{\lambda_1}} [\p_x + W(x)] \Psi_1(x) =
\frac{C}{\sqrt{k_1^2 - k_0^2}} 
[k_1 \cos(k_1 x) + k_0 \tan(k_0 x) \sin(k_1 x)], 
\end{equation}
which gives rise to the next superpotential
\begin{equation}
W'(x) = - \frac{\p_x \Psi_1'(x)}{\Psi_1'(x)} = 
\frac{(k_1^2 - k_0^2) \sin(k_1 x)}{k_1 \cos(k_1 x) + k_0 \tan(k_0 x) \sin(k_1 x)}
- k_0 \tan(k_0 x).
\end{equation}
Based on this, it is straightforward to work out the potential $V''(x)$ of the 
second descendant. However, the expression is not very illuminating, and we 
thus do not show it here.

It is instructive to consider special cases of the self-adjoint extension 
parameter $\gamma$, starting with the standard textbook case $\gamma = \infty$. 
The various potentials and the corresponding wave functions of the lowest 
energy eigenstates are illustrated for the original Hamiltonian $H$ and its
supersymmetric descendants $H'$ and $H''$ in figure 3.
\begin{figure}[t]
\begin{center}
\epsfig{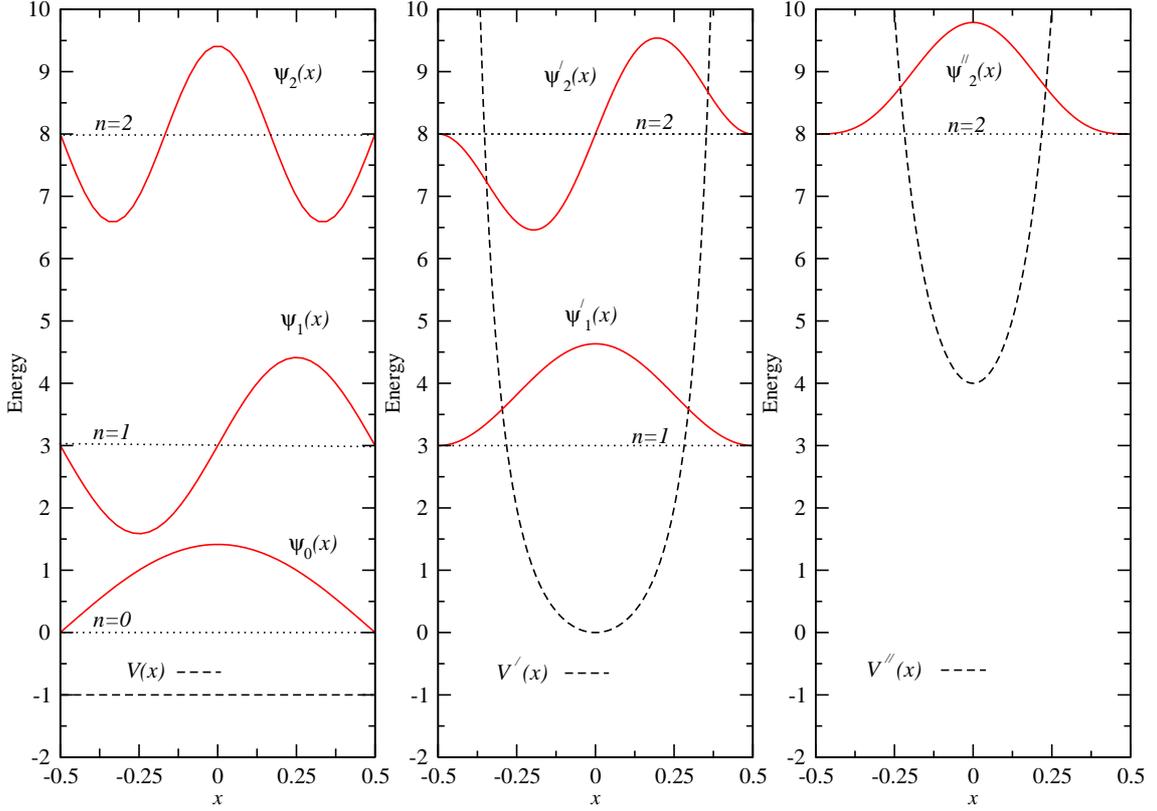}
\end{center}
\caption{\it A particle in a box with standard Dirichlet boundary conditions
(i.e.\ $\gamma = \infty$). The original system (left panel) is compared to its
first two supersymmetric descendants (middle and right panel). The corresponding
potentials $V(x)$, $V'(x)$, and $V''(x)$ are shown together with the low-energy
spectrum (horizontal dotted lines). The wave functions 
$\Psi_n(x)$, $\Psi_n'(x)$, and $\Psi_n''(x)$ are drawn using the corresponding 
energy level as the $x$-axis.}
\end{figure}
The $\gamma = 0$ case, which corresponds to Neumann boundary conditions, 
is illustrated in figure 4. In that case, the ground state has a constant wave 
function $\Psi_0(x) = \sqrt{1/L}$ of zero energy. Hence, the superpotential 
simply vanishes, i.e.\ $W(x) = 0$, and, in addition, $V(x) = V'(x) = 0$. Since 
the superpartner always obeys Dirichlet boundary conditions, i.e.\ 
$\gamma' = \infty$, in this case the first supersymmetric descendant coincides 
with the standard textbook case of a particle in a box.
\begin{figure}[t]
\begin{center}
\epsfig{file=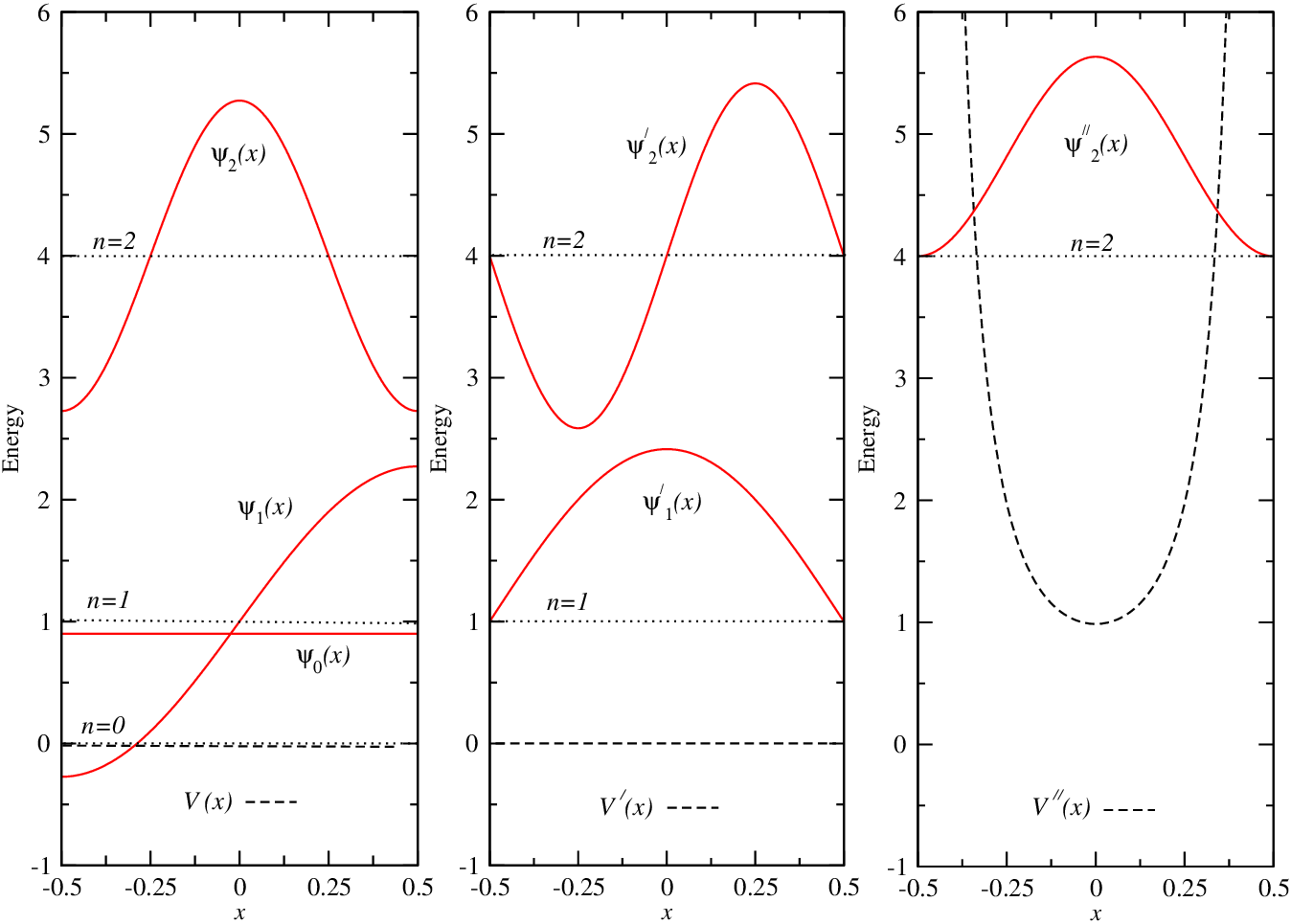,width=\textwidth}
\end{center}
\caption{\it A particle in a box with Neumann boundary conditions (i.e.\ 
$\gamma = 0$). The original system (left panel) is compared to its first two 
supersymmetric descendants (middle and right panel). The corresponding
potentials $V(x)$, $V'(x)$, and $V''(x)$ are shown together with the low-energy
spectrum (horizontal dotted lines). The wave functions 
$\Psi_n(x)$, $\Psi_n'(x)$, and $\Psi_n''(x)$ are drawn using the corresponding 
energy level as the $x$-axis.}
\end{figure}

For $\gamma < 0$, the ground state has negative energy, and is given by
\begin{equation}
\label{ground}
\Psi_0(x) = B \cosh(\varkappa_+ x), \ 
\frac{\gamma}{\varkappa_+} = - \tanh(\varkappa_+ L/2),
\end{equation}
such that the superpotential then takes the form
\begin{equation}
W(x) = - \frac{\p_x \Psi_0(x)}{\Psi_0(x)} = - \varkappa_+ \tanh(\varkappa_+ x).
\end{equation}
We thus obtain
\begin{eqnarray}
&&2 m V(x) = - \p_x W(x) + W(x)^2 = \varkappa_+^2, \nonumber \\
&&2 m V'(x) = \p_x W(x) + W(x)^2 =  
\varkappa_+^2  \left(1 - \frac{2}{\cosh^2(\varkappa_+ x)}\right).
\end{eqnarray}
For $- 2/L < \gamma < 0$, the first excited state has positive energy and is
still given by eq.(\ref{sin}). The ground state of the first supersymmetric 
descendant is then obtained as
\begin{equation}
\Psi_1'(x) = \frac{1}{\sqrt{\lambda_1}} [\p_x + W(x)] \Psi_1(x) =
\frac{C}{\sqrt{k_1^2 + \varkappa_+^2}} 
[k_1 \cos(k_1 x) - \varkappa_+ \tanh(\varkappa_+ x) \sin(k_1 x)], 
\end{equation}
which gives rise to the next superpotential
\begin{equation}
W'(x) = - \frac{\p_x \Psi_1'(x)}{\Psi_1'(x)} = 
\frac{(k_1^2 + \varkappa_+^2) \sin(k_1 x)}
{k_1 \cos(k_1 x) - \varkappa_+ \tanh(\varkappa_+ x) \sin(k_1 x)} + 
\varkappa_+ \tanh(\varkappa_+ x).
\end{equation}
The special case $\gamma = - 2/L$ is illustrated in figure 5. In that case, the 
first excited state, $\Psi_1(x) = \sqrt{12/L^3} x$, has zero energy.
\begin{figure}[t]
\begin{center}
\epsfig{file=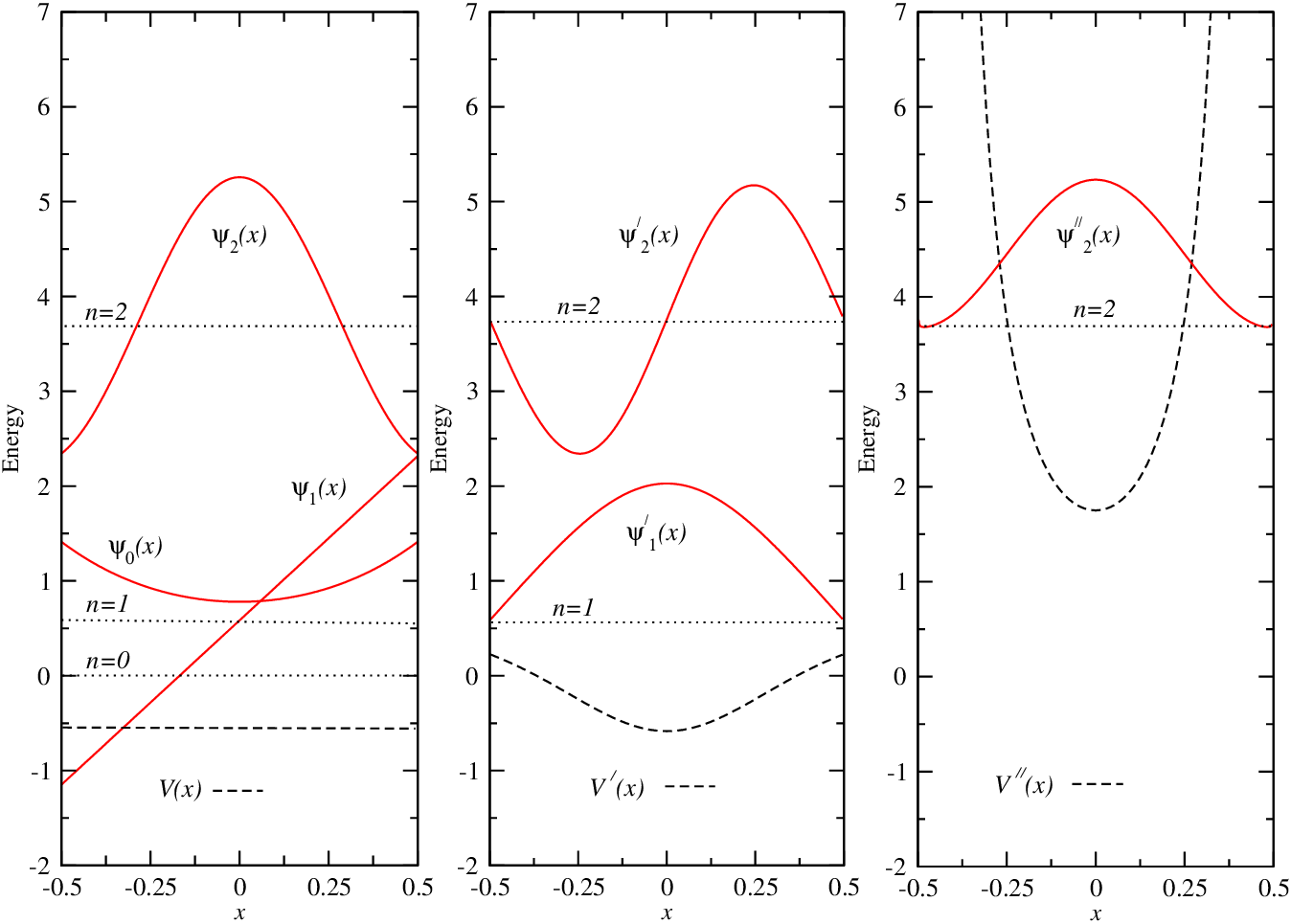,width=\textwidth}
\end{center}
\caption{\it A particle in a box with $\gamma = - 2/L$. The original system 
(left panel) is compared to its first two supersymmetric descendants (middle 
and right panel). The corresponding potentials $V(x)$, $V'(x)$, and $V''(x)$ 
are shown together with the low-energy spectrum (horizontal dotted lines). The 
wave functions $\Psi_n(x)$, $\Psi_n'(x)$, and $\Psi_n''(x)$ are 
drawn using the corresponding energy level as the $x$-axis.}
\end{figure}

For completeness, let us finally investigate the case $\gamma < - 2/L$. Then
both the ground state of eq.(\ref{ground}) and the first excited state have 
negative energy, and
\begin{equation}
\Psi_1(x) = C \sinh(\varkappa_- x), \ 
\frac{\gamma}{\varkappa_-} = - \coth(\varkappa_- L/2).
\end{equation}
The ground state of the first supersymmetric descendant is now given by
\begin{eqnarray}
\Psi_1'(x)&=&\frac{1}{\sqrt{\lambda_1}} [\p_x + W(x)] \Psi_1(x) \nonumber \\
&=&\frac{C}{\sqrt{\varkappa_+^2 - \varkappa_-^2}} 
[\varkappa_- \cosh(\varkappa_- x) - 
\varkappa_+ \tanh(\varkappa_+ x) \sinh(\varkappa_- x)], 
\end{eqnarray}
which gives rise to the next superpotential
\begin{equation}
W'(x) = - \frac{\p_x \Psi_1'(x)}{\Psi_1'(x)} = 
\frac{(\varkappa_+^2 - \varkappa_-^2) \sinh(\varkappa_- x)}
{\varkappa_- \cosh(\varkappa_- x) - 
\varkappa_+ \tanh(\varkappa_+ x) \sinh(\varkappa_- x)} + 
\varkappa_+ \tanh(\varkappa_+ x).
\end{equation}

\subsection{Particle in a Box with an Additional Point Interaction}

In this subsection, we consider a particle in a box with Robin boundary
conditions characterized by the self-adjoint extension parameter $\gamma$,
subject to an additional parity-invariant point interaction at $x_0 = 0$,
described by the self-adjoint extension parameters $a = \pm 1$ and $c$.

First, we consider positive energy states of even parity for which
\begin{equation}
\Psi_{k+}(x) = D_+ \cos(k|x| + \delta).
\end{equation}
It is straightforward to work out the equation for the corresponding energy 
values. For $a = 1$, one obtains
\begin{equation}
\tan\delta = - \frac{c}{2k}, \quad 
\frac{2 \gamma + c}{2 k - c \gamma/k} = \tan(kL/2),
\end{equation}
while for $a = -1$
\begin{equation}
\cos\delta = 0, \quad \frac{\gamma}{k} = - \cot(kL/2),
\end{equation}
Similarly, for the parity-odd states of positive energy one has
\begin{equation}
\Psi_{k-}(x) = D_- \sign(x) \cos(k|x| + \delta).
\end{equation}
In that case, for $a = 1$, one obtains
\begin{equation}
\cos\delta = 0, \quad \frac{\gamma}{k} = - \cot(kL/2),
\end{equation}
while for $a = -1$
\begin{equation}
\tan\delta = \frac{c}{2k}, \quad 
\frac{2 \gamma - c}{2 k + c \gamma/k} = \tan(kL/2),
\end{equation}
This shows that the energy spectrum of the even (odd) parity states for $a = 1$
and $c$ is the same as the one of the odd (even) parity states for $a = -1$ and
$- c$.

Let us also consider negative energy states, first with even parity
\begin{equation}
\Psi_+(x) = B_+ \exp(- \varkappa_+ |x|) + C_+ \exp(\varkappa_+ |x|).
\end{equation}
For $a = 1$, one then obtains
\begin{equation}
\frac{(\varkappa_+ - \gamma)(2 \varkappa_+ - c)}
{(\varkappa_+ + \gamma)(2 \varkappa_+ + c)} = \exp(\varkappa_+ L),
\end{equation}
while for $a = -1$
\begin{equation}
\frac{\gamma}{\varkappa_+} = - \tanh(\varkappa_+ L/2).
\end{equation}
Similarly, for the negative energy states with odd parity
\begin{equation}
\Psi_-(x) = \sign(x)[B_- \exp(- \varkappa_+ |x|) + C_- \exp(\varkappa_+ |x|)],
\end{equation}
with $a = 1$ one finds
\begin{equation}
\frac{\gamma}{\varkappa_-} = - \coth(\varkappa_- L/2),
\end{equation}
and with $a = -1$ one obtains
\begin{equation}
\frac{(\varkappa_- - \gamma)(2 \varkappa_- + c)}
{(\varkappa_- + \gamma)(2 \varkappa_- - c)} = \exp(\varkappa_- L).
\end{equation}
The corresponding energy spectrum is illustrated in figure 6, both for a
repulsive and for an attractive point interaction.
\begin{figure}[t]
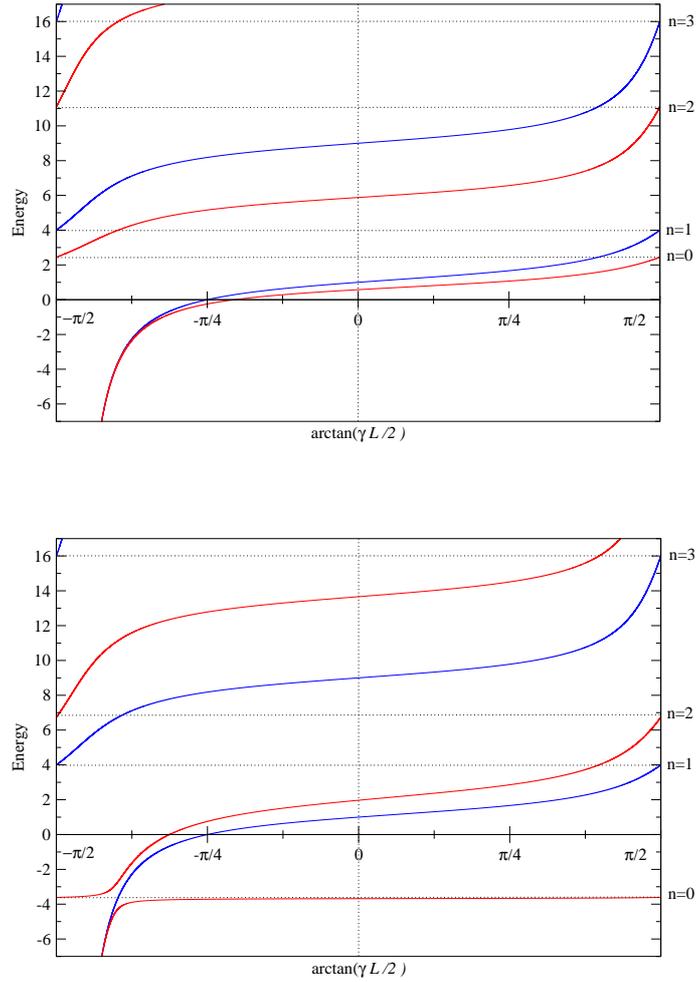

\begin{center}
\epsfig{file=SUSY_deltaGamma1.eps,width=0.6\textwidth} \\ \vskip1.2cm
\epsfig{file=SUSY_deltaGamma2.eps,width=0.6\textwidth}
\end{center}
\caption{\it  Energy spectrum of a particle in a box as a function
of the self-adjoint extension parameter $\gamma$ with a repulsive (top panel)
or attractive (bottom panel) point interaction with $c = \pm 12m$. The $x$-value
represents $\arctan(\gamma L/2) \in [- \tfrac{\pi}{2},\tfrac{\pi}{2}]$, which 
corresponds to $\gamma \in [-\infty,\infty]$. The $y$-value represents the 
energies $E_n$ (with $n = 0, 1, 2, 3$) in units of $\pi^2/(2 m L^2)$. The 
states with odd $n$ are unaffected by the point interaction and are identical
with those of a particle in a box without point interaction (figure 2).}
\end{figure}

\subsection{Bulk-Boundary Resonance and its Supersymmetric Descendant}

In the bottom panel of figure 6, at negative values of $\gamma$, one notices an 
avoided level crossing between the ground state and the second excited state.
Such avoided level crossings are characteristic of a resonance in a finite
volume \cite{Wie89,Lue91}. In this case, we encounter a bulk-boundary resonance,
with the particle partially localized at the walls, and partially at the center 
of the box, due to the attractive point interaction.  For definiteness, we set 
$a = 1$. The situation for $a = -1$ is analogous.

As a resonance condition, let us demand equal probability density at the walls 
and at the center, i.e.\ $|\Psi_+(\pm L/2)|^2 = |\Psi_+(0)|^2$. It is easy to 
convince oneself that this implies $c = 2 \gamma$, in which case the ground
state wave function reduces to
\begin{equation}
\Psi_0(x) = \Psi_+(x) = B \cosh[\varkappa_+ (|x| - L/4)], \
\frac{\gamma}{\varkappa_+} = - \tanh(\varkappa_+ L/4), \ \varkappa_+ > - \gamma.
\end{equation}
The first excited state takes the form
\begin{equation}
\Psi_1(x) = \Psi_-(x) = C \sinh(\varkappa_- x), \quad
\frac{\gamma}{\varkappa_-} = - \coth(\varkappa_- L/2),
\end{equation}
while the second excited state, which resonates with $\Psi_0(x)$, is given by
\begin{eqnarray}
&&\Psi_2(x) = \widetilde{\Psi}_+(x) = 
- D \sinh[\widetilde{\varkappa}_+ (|x| - L/4)], \nonumber \\
&&\frac{\gamma}{\widetilde{\varkappa}_+} = 
- \coth(\widetilde{\varkappa}_+ L/4), \ \widetilde{\varkappa}_+ < - \gamma.
\end{eqnarray}
These wave functions and their energies are illustrated in the left panel of
figure 7.
\begin{figure}[t]
\begin{center}
\epsfig{file=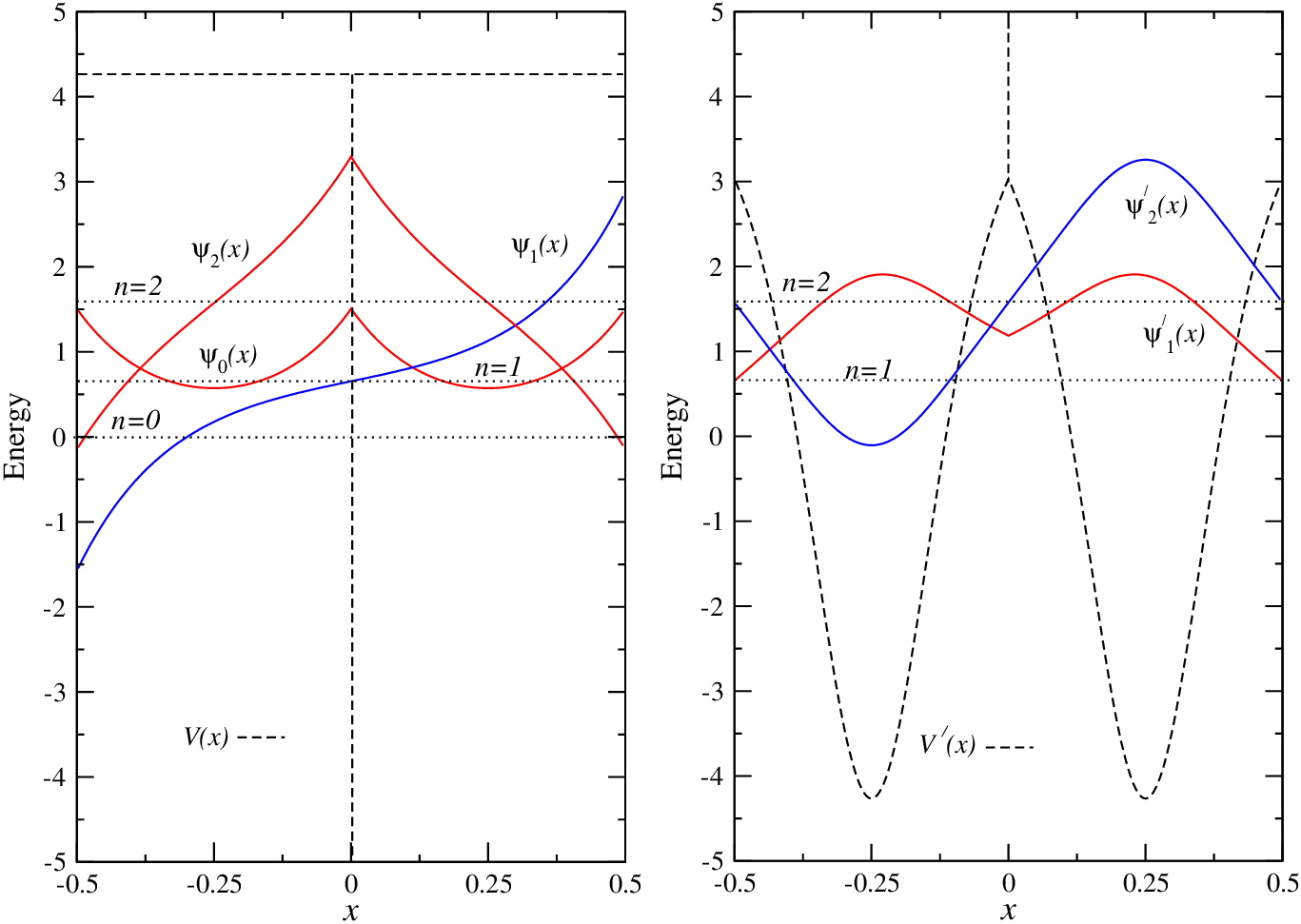,width=\textwidth}
\end{center}
\caption{\it  A bulk-boundary resonance of a particle in a box with an 
additional point interaction with $a = 1$. The original system (left panel) is 
compared to its supersymmetric descendant (right panel). The corresponding
potentials $V(x)$ and $V'(x)$ are shown together with the low-energy spectrum 
(horizontal dotted lines). The wave functions $\Psi_n(x)$ and $\Psi_n'(x)$ are 
drawn using the corresponding energy level as the $x$-axis. The downward and 
upward vertical dashed lines symbolize an attractive or repulsive point 
interaction, respectively.}
\end{figure}

Let us now construct the superpotential associated with the bulk-boundary
resonance
\begin{equation}
W(x) = - \frac{\p_x \Psi_0(x)}{\Psi_0(x)} = 
- \varkappa_+ \sign(x) \tanh[\varkappa_+ (|x| - L/4)]. 
\end{equation}
The original potential and that of the superpartner then take the form
\begin{eqnarray}
&&2 m V(x) = - \p_x W(x) + W(x)^2 = \varkappa_+^2 \nonumber \\
&&2 m V'(x) = \p_x W(x) + W(x)^2 = 
\varkappa_+^2 \left(1 - \frac{2}{\cosh^2[\varkappa_+ (|x| - L/4)]}\right). 
\end{eqnarray}
Interestingly, $V'(x)$ represents a double-well potential, with a repulsive
point interaction at $x_0 = 0$, characterized by 
\begin{equation}
c' = - c = - 2 \gamma = 2 \varkappa_+ \tanh(\varkappa_+ L/4). 
\end{equation}
The ground and first excited state of the supersymmetric descendant are given by
\begin{eqnarray}
\Psi_1'(x)\!\!&=&\!\!\frac{1}{\sqrt{\lambda_1}} [\p_x + W(x)] \Psi_1(x) 
\nonumber \\
&=&\!\!\frac{C}{\sqrt{\varkappa_+^2 - \varkappa_-^2}}
\{\varkappa_- \cosh(\varkappa_- x) - 
\sign(x) \varkappa_+ \tanh[\varkappa_+ (|x| - L/4)] \sinh(\varkappa_- x)\},
\nonumber \\
\Psi_2'(x)\!\!&=&\!\!\frac{1}{\sqrt{\lambda_2}} [\p_x + W(x)] \Psi_2(x) 
\nonumber \\
&=&\!\!\frac{D \sign(x)}{\sqrt{\varkappa_+^2 - \widetilde{\varkappa}_+^2}}
\{\widetilde{\varkappa}_+ \cosh[\widetilde{\varkappa}_+ (|x| - L/4)]
\nonumber \\
&-&\!\!\varkappa_+ \tanh[\varkappa_+ (|x| - L/4)] 
\sinh[\widetilde{\varkappa}_+ (|x| - L/4)]\}. 
\end{eqnarray}
These states together with the corresponding potential $V'(x)$ are illustrated
in the right panel of figure 7. 

In the original system, the first excited state is unaffected by the point 
interaction and is identical with the one of just the particle in the box. The
ground state and the second excited state, on the other hand, resonate with one
another and are both localized on the walls as well as on the puncture at 
$x_0 = 0$. In the spectrum, the resonance manifests itself by an avoided level 
crossing. When one proceeds to the supersymmetric descendant, the ground state 
is removed and the first excited state of the original system turns into the
ground state of the superpartner. Interestingly, while this state was 
unaffected by the attractive point interaction of the original system, it is 
affected by the repulsive point interaction of the supersymmetric descendant.
Similarly, the second excited state of the original system, which was affected
by the attractive point interaction, turns into the first excited state of the
superpartner, but is now unaffected by its repulsive point interaction. What has
become of the resonance of the two states, now that the original ground state 
has been removed from the supersymmetric descendant? As we see from the right
panel of figure 7, the superpartner has a double-well potential, and its ground
and first excited states are almost degenerate, with a splitting due to 
tunneling processes between the two wells. Indeed, the regions near the walls
and near the puncture, which were energetically favored in the original system,
are disfavored in the supersymmetric descendant. This shows how one and the same
spectrum (except for the ground state $\Psi_0(x)$) can arise from quite 
different physical phenomena, in one case a bulk-boundary resonance, in the 
other case tunneling in a double-well potential.

The corresponding situation for $a = -1$ is illustrated in figure 8, which
confirms that the spectrum is the same as for $a = 1$, but even and odd parity
states exchange their roles. In particular, the ground state is now parity-odd,
and the first excited state is parity-even. 
\begin{figure}[t]
\begin{center}
\epsfig{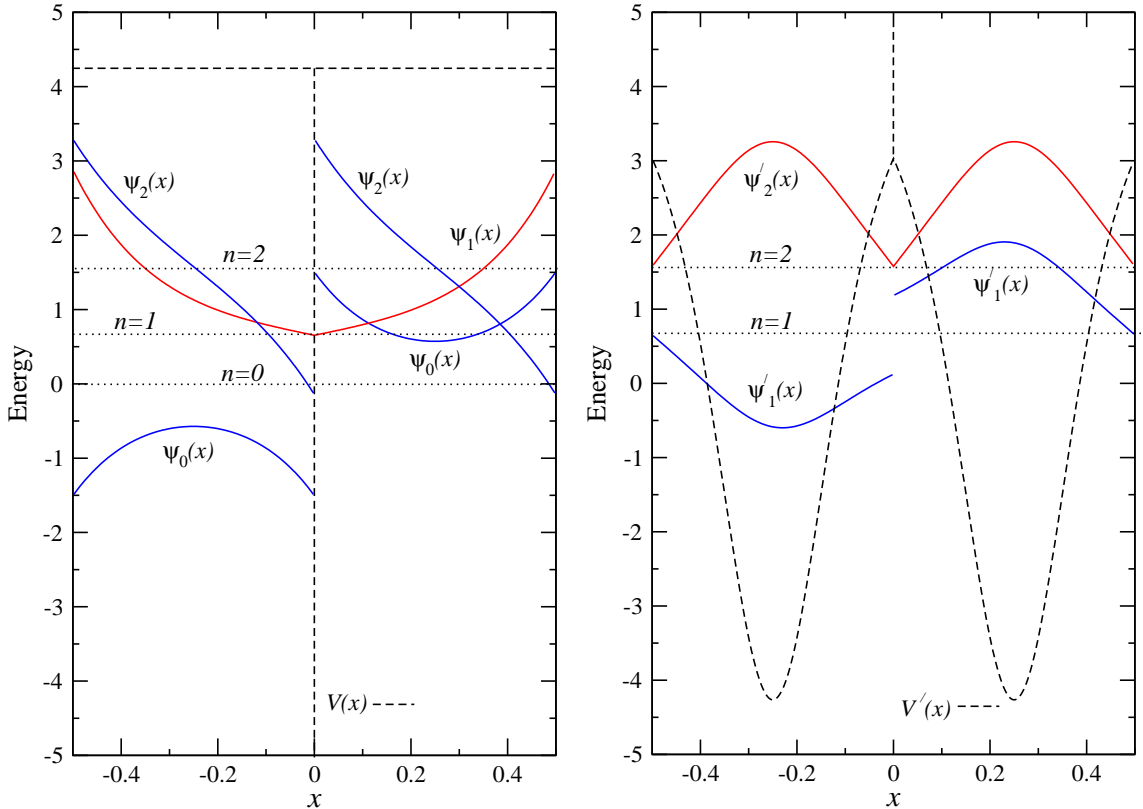} 
\end{center}
\caption{\it   A bulk-boundary resonance of a particle in a box with an 
additional point interaction with $a = -1$. The original system (left panel) is 
compared to its supersymmetric descendant (right panel). The corresponding
potentials $V(x)$ and $V'(x)$ are shown together with the low-energy spectrum 
(horizontal dotted lines). The wave functions $\Psi_n(x)$ and $\Psi_n'(x)$ are 
drawn using the corresponding energy level as the $x$-axis. The downward and 
upward vertical dashed lines symbolize an attractive or repulsive point 
interaction, respectively.}
\end{figure}

\section{Conclusions}

We have investigated the supersymmetric descendants of self-adjointly extended
Hamiltonians. The infinite-wall boundary condition of a particle on a half-line
is characterized by a family of self-adjoint extensions, parameterized by
$\gamma \in \R$. Interestingly, all corresponding supersymmetric descendants 
have $\gamma' = \infty$ and thus obey standard Dirichlet boundary conditions.
A particle on a punctured line with a point interaction at the puncture $x_0$
is characterized by a 4-parameter family of self-adjoint extensions. Remarkably,
in that case, the corresponding supersymmetric descendants are not automatically
self-adjoint. Indeed, only a 3-parameter sub-family of Hamiltonians has 
supersymmetric descendants which are themselves self-adjoint. This sub-family
is characterized by the continuity of the probability density at the puncture.

We have also constructed the self-adjoint extensions of the operator 
${\cal Q}_+$ constructed from the supercharge. They form a 1-parameter family 
on the half-line, and a 4-parameter family on the punctured line. Yet, only one 
specific value of the self-adjoint extension parameter (namely $\eta = 0$) is 
physical on the half-line, and only a 2-parameter sub-class is physical on the 
punctured line. This is because we have considered $H$ and $H'$ as two distinct 
physical systems, and not as two parts of a bigger system. While there is a 
1-parameter family of self-adjoint extensions of $H$ on the half-line 
(parameterized by $\gamma$), there is no remaining self-adjoint extension 
parameter in ${\cal Q}_+$, after we put $\eta = 0$. This is because the 
information on $\gamma$ is encoded in the superpotential. Similarly, on the 
punctured line there is a 3-parameter family of self-adjoint extensions of $H$ 
(parameterized by $a$, $c$, and $\theta$), for which the supersymmetric 
descendant $H'$ is also self-adjoint. At the same time, there is only a 
2-parameter family of self-adjoint extensions of ${\cal Q}_+$ (parameterized by 
$a$ and $\theta$). This is again because the information about the third 
parameter $c$ is encoded in the superpotential. This clarifies the relations 
between the self-adjoint extensions of $H$, $H'$, and ${\cal Q}_+$.

We have also examined concrete problems of a particle in a box, with or without
an additional point interaction. Among other things, we found that the standard
textbook problem of a particle in a box with Dirichlet boundary conditions is
itself a supersymmetric descendant. Its supersymmetric precursor is the 
corresponding problem with Neumann boundary conditions. Robin boundary 
conditions with $\gamma < 0$ give rise to negative energy states localized at
the walls. Such boundary states can resonate with states localized in the bulk,
which gives rise to an avoided level crossing in a finite volume. We have
investigated the supersymmetric descendant of such a resonance and found that 
it corresponds to two almost degenerate states in a double-well potential.

By applying self-adjoint extensions to supersymmetric quantum mechanics, we 
have extended the set of exactly solvable quantum mechanics problems.
Self-adjoint extensions are not just a mathematical curiosity, but have great
physical relevance. The self-adjoint extension parameters just characterize the
low-energy features of an idealized boundary, such as an impenetrable infinite 
energy barrier, or an ultra-short-range attractive potential in a tiny region of
space. Using the theory of self-adjoint extensions greatly simplifies the
modeling of such situations. In fact, some of the calculations performed here 
are so simple that they could easily be incorporated into the teaching of 
quantum mechanics. We conclude this paper by expressing our hope that, in the
future, the powerful theory of self-adjoint extensions may make a stronger 
appearance in textbooks and in the teaching of quantum mechanics.

\section*{Acknowledgments}

This publication was made possible by NPRP grant \# NPRP 5 - 261-1-054 from
the Qatar National Research Fund (a member of the Qatar Foundation). The
statements made herein are solely the responsibility of the authors.

\end{document}